\documentclass[epj]{svjour}
\usepackage{graphics}
\newcommand{\be}{\begin{equation}}
\newcommand{\ee}{\end{equation}}
\newcommand{\bea}{\begin{eqnarray}}
\newcommand{\eea}{\end{eqnarray}}
\begin{document}
\title{Nuclear response in a finite-temperature relativistic framework}
\author{ Elena Litvinova\inst{1}$^,$\inst{2} \and  Herlik Wibowo\inst{1}% etc
}                     % Do not remove
\institute{Department of Physics, Western Michigan University, Kalamazoo, MI 49008, USA \and National Superconducting Cyclotron Laboratory, Michigan State University, East Lansing, MI 48824, USA}
\date{Received: date / Revised version: date}
\abstract{A thermal extension of the relativistic nuclear field theory is formulated for the nuclear response.  
The Bethe-Salpeter equation (BSE) with the time-dependent kernel for the particle-hole response is treated within the Matsubara Green's function formalism. We show that, with the help of a temperature-dependent projection operator on the subspace of the imaginary time (time blocking), it is possible to reduce the BSE for the nuclear response function to a single frequency variable equation also at finite temperature. The approach is implemented self-consistently in the framework of quantum hadrodynamics based on the meson-nucleon Lagrangian. The method is applied to the monopole, dipole and quadrupole response of $^{48}$Ca and to the dipole response of the tin isotopes $^{100,120,132}$Sn, in particular, to a study of the evolution of nuclear collective oscillations with temperature. The article is dedicated to the memory of Pier Francesco Bortignon and devoted to the developments related to his pioneering ideas.
%Insert your abstract here.
%
\PACS{\  21.10.-k, 21.30.Fe, 21.60.-n, 24.10.Cn, 24.30.Cz
%      {PACS-key}{describing text of that key}   \and
%      {PACS-key}{describing text of that key}
     } % end of PACS codes
} %end of abstract
\maketitle
\section{Introduction}
\label{intro}

Response to external probes is one of the major characteristics of strongly-correlated systems, which provides complete information about them. In nuclear systems, the spectral properties extracted from various nuclear reactions serve also for understanding the underlying nuclear forces. Being extremely complicated already for a two-nucleon system, the nucleon-nucleon interaction is, in addition, considerably modified by numerous processes in the strongly-correlated medium, which are commonly called medium polarization. It was recognized early by A. Bohr
and B.R. Mottelson \cite{BohrMottelson1969,BohrMottelson1975} that the major contribution to the medium polarization originates from the coupling between nucleonic and vibrational degrees of freedom, at least
in medium-mass and heavy nuclei. Since then, this idea has advanced many directions in both theoretical and experimental nuclear physics. Pier Francesco Bortignon was one of the pioneers of the nuclear field theory (NFT) which developed the concept of the interaction between fermionic and  bosonic (vibrational) degrees of freedom in atomic nuclei \cite{Broglia1976,BortignonBrogliaBesEtAl1977,Bortignon1978,Bortignon1981,BertschBortignonBroglia1983,MahauxBortignonBrogliaEtAl1985,Bortignon1986,Bortignon1997,ColoBortignon2001}. These developments lead to a very successful description of a large amount of experimental data and advanced tremendously the understanding of many nuclear structure phenomena, especially those related to collective effects.  They also inspired experimental groups in various fields of research, such as giant resonances at finite temperature and multi-phonon excitations.
This has become possible after a
number of works on single-particle and collective excitations in hot nuclei within the thermal NFT  \cite{Bortignon1986,BrogliaBortignonBracco1992,Giovanardi1996,Donati1996,Giovanardi1998} as well as on the theory of thermal shape fluctuations \cite{OrmandBortignonBrogliaEtAl1990,OrmandBortignonBroglia1996}. A large corpus of theoretical knowledge and experimental data was summarized in the volume "Giant Resonances: Nuclear Structure at Finite Temperature" by Pier Francesco Bortignon and his co-authors Angela Bracco and Ricardo A. Broglia  \cite{Bortignon1998}.
In his late years, he was involved in new advancements of the NFT \cite{Bortignon2016,Bortignon2016a,Broglia2016a,Colo2017}, which also open new horizons for future research.

%%%%%%%
In this article dedicated to his memory we would like to focus on one of our recent developments which are closely related to the pioneering ideas of Pier Francesco Bortignon on the nuclear response at finite temperature. Inspired by those,
the temperature dependence of the nuclear response, mostly in the dipole channel and mostly of its high-energy part associated with the giant dipole resonance (GDR), was extensively studied experimentally in the past \cite{Gaardhoje1984a,GaardhojeEllegaardHerskindEtAl1986,Bracco1989,Ramakrishnan1996,Mattiuzzi1997,Heckman2003}, see also a relatively recent review \cite{Santonocito2006} and a newer study of Refs. \cite{Wieland2006,Bracco2007}. In more recent experiments on the dipole response, a concentration of strength has been observed in the low-energy region \cite{SavranAumannZilges2013,Isaak2019}, being most prominent in neutron-rich nuclei. Its origin was attributed to the coherent oscillation of the neutron skin around the isospin-saturated core, which is called pygmy dipole resonance (PDR). In the gamma decay of quasicontinuum excitations another enhancement of the dipole strength has been found at very low gamma energies \cite{VoinovAlginAgvaanluvsanEtAl2004,WiedekingBernsteinKrtickaEtAl2012}. These new features of the low-energy strength functions become especially interesting in the context of astrophysical modeling, because they can, in particular, significantly enhance the  reaction rates of the rapid neutron-capture nucleosynthesis (r-process) \cite{MumpowerSurmanMcLaughlinEtAl2016} and modify the nuclear matter equation of state. 
%%%%%
Similar enhancements of the low-lying spin-isospin response are found in the theoretical finite-temperature studies of Refs. \cite{Minato2009,Dzhioev2010,Niu2011}, which indicate that the electron capture and beta decay rates in the astrophysical environments are also affected by the sensitivity of nuclear transitions to finite temperature. Thus, both phases of the r-process (neutron capture and beta decay), which occurs in the neutron star mergers (NSM), and electron capture, which is crucial for the core-collapse supernovae (CCSN) simulations, require precise knowledge about low-energy nuclear transitions at finite temperature. The nature and progress of the r-process depend on the specific astrophysical scenario, which can vary in rather broad limits, however, a consistent description of merging neutron stars require the information about the CCSN which precedes the neutron star formation. As discussed in Ref. \cite{Aprahamian2018}, ultimately, identical simulation frameworks are to be used for both the NSM and the CCSN.

As the complete description of the r-process requires information about many nuclei which are beyond current experimental capabilities, a reliable theory of the finite-temperature nuclear response is highly desirable. 
An accurate theory for the response of nuclei at finite temperature, or compound nuclei, which is required for applications, is very challenging. 
The most common practice for the microscopic approaches to the finite-temperature nuclear response 
is to confine the framework by the simplest one-loop approximation represented by the thermal random phase approximation (TRPA) \cite{RingRobledoEgidoEtAl1984,NiuPaarVretenarEtAl2009}, thermal quasiparticle RPA (TQRPA) \cite{RingRobledoEgidoEtAl1984,YuekselColoKhanEtAl2017}, and by the continuum TQRPA \cite{LitvinovaKamerdzhievTselyaev2003,KhanVanGiaiGrasso2004,LitvinovaBelov2013}. Some extensions beyond T(Q)RPA have been proposed over the years,  for instance, the finite- temperature self-consistent RPA \cite{DukelskyRoepkeSchuck1998},  the collision-integral approach 
%with incoherent two-particle two-hole (2p2h) excitations
\cite{LacroixChomazAyik1998}, the finite-temperature second RPA (SRPA) \cite{AdachiSchuck1989,Yannouleas1986}, the thermal NFT with a coupling of nucleons to collective surface vibrations \cite{Bortignon1986}, and the quasiparticle-phonon model formulated in terms of the thermofield dynamics \cite{Storozhenko2004}. These microscopic approaches elaborated on different aspects of the finite-temperature nuclear response and, together with the theories of thermal shape fluctuations \cite{Kusnezov1998,OrmandBortignonBroglia1996} advanced considerably the understanding of spectral properties of hot nuclei.

In this article we discuss another microscopic approach to the finite-temperature nuclear response, which was developed recently in a relativistic framework. The aim of this approach is to advance the relativistic version of the nuclear field theory (RNFT) 
\cite{LitvinovaRing2006,LitvinovaRingTselyaev2007,LitvinovaRingTselyaev2008,LitvinovaRingTselyaev2013,Litvinova2015,Litvinova2016,RobinLitvinova2016,RobinLitvinova2018} to the finite- temperature case. The RNFT, which is based on the covariant energy density functional of quantum hadrodynamics, extends the relativistic RPA (RRPA) by the (quasi)particle-vibration coupling in a parameter-free way. This approach performs very well in the description of nuclear transitions from ground to excited states 
\cite{LitvinovaRingTselyaevEtAl2009,LitvinovaLoensLangankeEtAl2009,LitvinovaRingTselyaev2010,EndresLitvinovaSavranEtAl2010,TamiiPoltoratskaNeumannEtAl2011,%
MassarczykSchwengnerDoenauEtAl2012,LitvinovaRingTselyaev2013,SavranAumannZilges2013,LanzaVitturiLitvinovaEtAl2014,PoltoratskaFearickKrumbholzEtAl2014,Oezel-TashenovEndersLenskeEtAl2014,EgorovaLitvinova2016}. It has been extended recently in Refs. \cite{LitvinovaWibowo2018,WibowoLitvinova2018} to finite temperatures and applied to the dipole response of selected medium-light and medium-heavy nuclei. In the present article we discuss some more details of the dipole response as well as further aspects of the finite-temperature theory, such as the monopole and quadrupole excitations and its broader impacts on the r-process and on the nuclear matter equation of state.  We will consider both the low-temperature range (1-2 MeV) relevant for the astrophysical applications, such as NSM and CCSN, and higher temperatures which can be achieved in heavy-ion collisions.

\section{Method}
\label{sec:1}
%and \cite{RefJ}
%\subsection{Subsection title}
%\label{sec:2}
%as required. Don't forget to give each section
%and subsection a unique label (see Sect.~\ref{sec:1}).
%
We begin with modeling a compound nucleus with the finite-temperature relativistic mean-field (RMF)
theory. The grand potential $\Omega(\lambda,T)$ %\cite{Sommermann1983} 
\be
\Omega(\lambda,T) = E  - T S - \lambda N
\label{Omega}
\ee
is minimized with the Lagrange multipliers $\lambda$ and $T$ determined by the
average energy $E$, particle number $N$, and the entropy $S$ \cite{Sommermann1983}. The 
quantities $N$ and $S$ are thermal averages with the 
one-body nucleonic density operator $\hat{\rho}$ :
\be
S = -k\mbox{Tr}({\hat{\rho}}\mbox{ln}\hat{\rho}), \ \ \ 
%\ E = Tr(\rho{\cal H}), \ \ \ 
\ N= \mbox{Tr}(\hat{\rho} {\hat{\cal N}}).
\label{SN}
\ee 
Here $\hat{\cal N}$ is the particle number operator, $k$ is the Boltzmann constant, and
the energy $E$ is a covariant functional of the nucleonic density and classical meson and photon fields $\phi_m$ \cite{VretenarAfanasjevLalazissisEtAl2005}:
\bea
E[\hat{\rho},\phi_m] &=& \mbox{Tr}[({\vec\alpha}\cdot{\vec p} + \beta M)\hat{\rho}] + \sum\limits_m\Bigl\{\mbox{Tr}[(\beta\Gamma_m\phi_m)\hat{\rho}] \pm \nonumber \\
&\pm& \int d^3r \Bigl[\frac{1}{2} ({\vec\nabla}\phi_m)^2 + U(\phi_m)\Bigr]\Bigr\}.
\label{cedf}
\eea
%${\cal H}$ is the nuclear Hamiltonian, 
The energy is thus determined by the nucleon mass $M$ and non-linear sigma-meson potentials $U(\phi_m)$ containing meson masses and meson-nucleon coupling vertices adjusted to bulk nuclear characteristics  \cite{Lalazissis1997,VretenarAfanasjevLalazissisEtAl2005}. In Eq. (\ref{cedf}) the sign "+" is associated with the scalar $\sigma$-meson, "-" with the vector $\omega$- and $\rho$-mesons, and the index "$m$" runs over the mesons, photon and Lorentz indices \cite{VretenarAfanasjevLalazissisEtAl2005}.
The variation of Eq. (\ref{Omega})
determines the operator of the nucleonic density with the 
unity trace 
\be
{\hat\rho} = \frac{e^{-({\hat{\cal H}}-\lambda{\hat{\cal N}})/kT }}{\mbox{Tr}[e^{-({\hat{\cal H}}-\lambda{\hat{\cal N}})/kT }]},   
\ee
where ${\hat{\cal H}} = \delta E[{\hat\rho},\phi_m]/\delta {\hat\rho} $. The 
eigenvalues 
of the density matrix are 
the Fermi-Dirac occupation factors:
\be
n_1(T) = n(\varepsilon_1, T) = \frac{1}{1 + \mbox{exp}\{\varepsilon_1 /T \}},
\label{fdo}
\ee
where the subscript "1" runs over the complete set of the single-particle quantum numbers in the Dirac-Hartree basis, which diagonalizes ${\hat{\cal H}}$ with the eigenvalues of 
the single-particle energies $\varepsilon_1 = {\tilde\varepsilon_1} - \lambda$ measured from the chemical potential $\lambda$. In Eq. (\ref{fdo}) and in the following we adopt the units where the value of the Boltzmann constant is $k =1$.
In the following, we will deal also with the counterpart of Eq. (\ref{fdo}) in the bosonic sector:
\be
%n_i = 
N_{\mu}(T) \equiv N(\Omega_{\mu}, T) = \frac{1}{\mbox{exp}\{\Omega_{\mu}/T\} - 1}, 
\ee
where $\Omega_{\mu}$ are the frequencies of the phonon modes  composed of correlated particle-hole excitations, which possess the quantum numbers $\mu$.

As in Refs. \cite{LitvinovaWibowo2018,WibowoLitvinova2018},
in this work we will consider non-superfluid nuclear systems, such as doubly-magic nuclei and nuclei at temperatures above the critical temperature, when pairing correlations vanish. In this case, the particle-hole finite-temperature four-point response function ${\mathcal{R}}(12,34)$ can be conveniently formulated in the Dirac-Hartree basis of the single-particle states defined above and in the imaginary-time representation, so that the number multi-indices $1 = \{k_1,\tau_1\}$ with the $k_1$ characterizing the single-nucleon quantum numbers and $\tau_1$ being the imaginary time. The response function ${\mathcal{R}}(12,34)$ obeys the Bethe-Salpeter equation (BSE)
\begin{eqnarray}
\label{BSE}
{\mathcal{R}}(12,34)={\widetilde{\mathcal{R}}}^{0}(12,34)+ \nonumber \\
+ \sum_{5678}^{\tau}{{\widetilde\mathcal{R}}}^{0}(12,56)\Big[{\widetilde V}(56,78) + {\mathcal W}(56,78)\Big]{\mathcal{R}}(78,34),
\end{eqnarray}
which is formally similar to the one at zero temperature, but in the imaginary-time representation. 
%In the following we will put $\tau_1=\tau_3, \tau_2=\tau_4$. 
In Eq. (\ref{BSE}) the summations imply integrations over the time arguments:
%\begin{equation}
%\widetilde{\mathcal{R}}^{0}(12,34)=-\widetilde{\mathcal{G}}(3,1)\widetilde{\mathcal{G}}(2,4)
%\end{equation}
%and
\begin{equation}
\sum^{\tau}_{12..}=\sum_{k_{1}k_{2}...}\int_{0}^{1/T}d\tau_{1}d\tau_{2}\cdots
\end{equation}
and the uncorrelated particle-hole response ${\widetilde\mathcal{R}}^{0}(12,34)= \\ {\widetilde\mathcal{G}}(3,1){\widetilde\mathcal{G}}(2,4)$ is defined via the one-body Matsubara Green's function ${\widetilde\mathcal{G}}$ in the thermal mean-field \cite{Abrikosov1965}:
\begin{eqnarray}
\label{mfgf}\widetilde{\mathcal{G}}(2,1)=\sum_{\sigma}\widetilde{\mathcal{G}}^{\sigma}(2,1),\\
\label{mfgf1}\widetilde{\mathcal{G}}^{\sigma}(2,1)=-\sigma\delta_{12}n[-\sigma\varepsilon_{1},T]
e^{-\varepsilon_{1}\tau_{21}}\theta(\sigma\tau_{21}),
\end{eqnarray}
where $\tau_{21}=\tau_{2}-\tau_{1}$ ($-1/T<\tau_{21}<1/T$), $\theta(\tau)$ is the Heaviside step-function and the index $\sigma=+1(-1)$ denotes the retarded (advanced) component of $\widetilde{\mathcal{G}}$.
%\begin{equation}
%\label{Fermi Occupation Number}n_{k_{1}}(\varepsilon_{k_{1}}-\mu,T):=n_{k_{1}}(\varepsilon_{k_{1}},T)=\frac{1}{e^{(\varepsilon_{k_{1}}-\mu)/T}+1}.
%\end{equation}
The Fourier transformation of Eqs. (\ref{mfgf},\ref{mfgf1}) with respect to the imaginary time gives the mean-field Matsubara Green's function in the domain of the discrete
imaginary energy variable:
\begin{eqnarray}
\widetilde{\mathcal{G}}_{21}(\varepsilon_{\ell}) = \frac{1}{2}\int^{1/T}_{-1/T}d\tau_{21}e^{i\varepsilon_{\ell}\tau_{21}}\widetilde{\mathcal{G}}(2,1)  
\nonumber \\
= \frac{\delta_{12}}{i\varepsilon_{\ell}-\varepsilon_{1}+\mu} = \delta_{12}\widetilde{\mathcal{G}}_{1}(\varepsilon_{\ell}),
\end{eqnarray} 
where $\varepsilon_{\ell}=(2\ell+1)\pi T$ with $\ell$ being integer numbers.

\begin{figure}
\resizebox{0.47\textwidth}{!}{\includegraphics{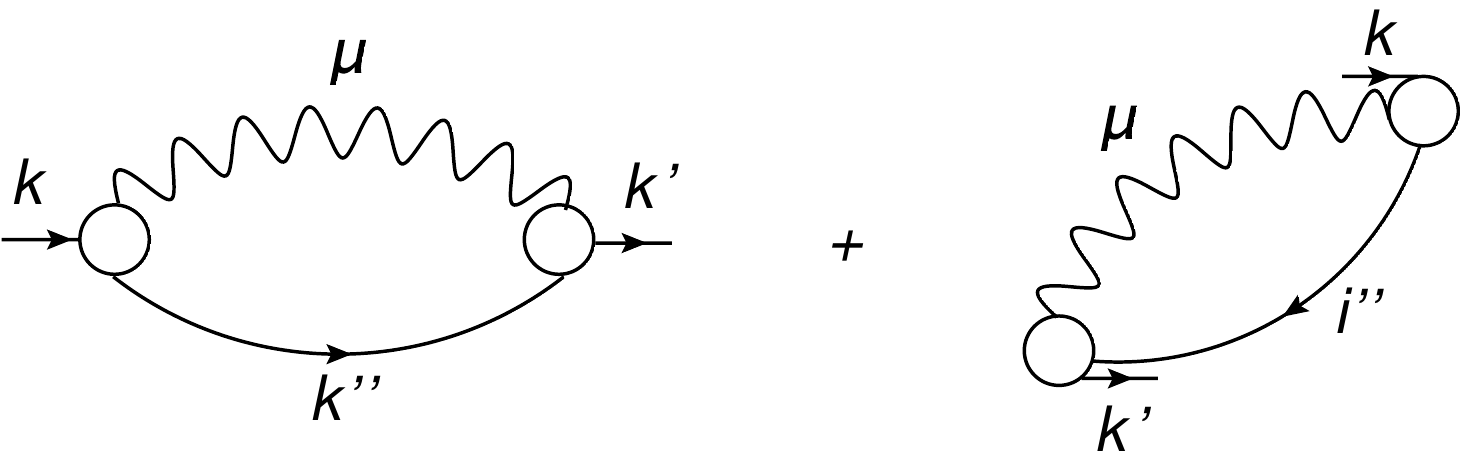}}
\caption{Diagrammatic representation of the self-energy $\Sigma^{e}$, the main building block of the particle-vibration coupling contributions. Straight lines represent one-fermion propagators, circles denote the particle-phonon coupling vertices, and the wiggly lines correspond to the phonon propagators.}
\label{fig:1}
\end{figure}

The interaction kernel of the BSE (\ref{BSE}) is divided into two parts $\widetilde V$ and $\cal W$, which are responsible for the short and long-range correlations, respectively. This division is very general and can be obtained, for instance, by the equation of motion method for a model-independent Hamiltonian with a time-independent two-body bare interaction \cite{AdachiSchuck1989,DukelskyRoepkeSchuck1998,SchuckTohyama2016}. As in our previous works \cite{LitvinovaRingTselyaev2007,LitvinovaRingTselyaev2008,LitvinovaRingTselyaev2010,LitvinovaRingTselyaev2013}, the short-range contribution $\widetilde V$ is associated with the exchange by effective mesons and photon parametrized in accordance with Eq. (\ref{cedf}) \cite{Lalazissis1997}, and the long-range term $\cal W$ is induced by the medium polarization effects. In practice, the short-range term is almost instantaneous, and its temporal locality translates to the energy independence of its Fourier transform. In case of neglecting the second long-range term, where the retardation effects are important, the BSE reduces to the response-theory form of the random phase approximation (RPA). If the latter term is retained, it translates to an energy-dependent kernel while the time integrations transform to the integrations over the energy variables of the internal fermionic loops. Because of a clear separation of these two terms, it is convenient to divide the problem of solving Eq. (\ref{BSE}) into two parts, namely: 
(i) to calculate the correlated propagator $\mathcal{R}^{e}$ from the equation (in the operator form):
\begin{equation}
\label{Operator Form of Re}\mathcal{R}^{e}={\widetilde{\mathcal R}}^{0}+{\widetilde{\mathcal R}}^{0}\mathcal{W}\mathcal{R}^{e}
\end{equation}
and (ii) to solve the remaining equation
\begin{equation}
\mathcal{R}=\mathcal{R}^{e}+\mathcal{R}^{e}\widetilde{V}\mathcal{R}
\label{FullResp}
\end{equation} 
for obtaining the full response $\mathcal{R}$. The problem of finding ${\mathcal R}^{e}$ is, thus, the central problem of accounting for the complex nuclear long-range correlations. The exact solution of this problem is hardly possible at both finite and zero temperatures and, thus, requires some model assumptions and simplifications which, at the same time, would retain the leading contribution of correlations beyond RPA. A truncation scheme based on retaining only two-body correlation functions leads to the following leading approximation for the long-range part of the interaction kernel:
\bea
{\cal W}(12,34) = \widetilde{\mathcal{G}}^{-1}(3,1)\Sigma^{e}(2,4)&+&\Sigma^{e}(3,1)\widetilde{\mathcal{G}}^{-1}(2,4) +\nonumber\\
&+& \mathcal{U}^{e}(12,34),
\eea 
where $\Sigma^{e}(3,1)$ and $\mathcal{U}^{e}(12,34)$ are the time-dependent parts of the fermionic self-energy and induced interaction, respectively. The diagrammatic expression of $\Sigma^{e}(3,1)$ is given in Fig. \ref{fig:1} and the induced interaction $\mathcal{U}^{e}(12,34)$ is related to it by the dynamical consistency condition 
\begin{eqnarray}
\Sigma^{e}_{12}(\varepsilon_{\ell}+\omega_{n})&-&\Sigma^{e}_{12}(\varepsilon_{\ell})\nonumber\\
&=&T\sum_{34}\sum_{\ell'}\mathcal{U}^{e}_{21,43}(\omega_{n},\varepsilon_{\ell},\varepsilon_{\ell'})\times\nonumber\\
&\times&[\widetilde{\mathcal{G}}_{34}(\varepsilon_{\ell'}+\omega_{n})-\widetilde{\mathcal{G}}_{34}(\varepsilon_{\ell'})],
\end{eqnarray}
which we assume as a finite-temperature generalization of the $T=0$ one \cite{KamerdzhievTertychnyiTselyaev1997} using
$\omega_{n}=2n\pi T$ with integer $n$.

Thus, the equation for the correlated propagator takes the form:
\begin{eqnarray}
\label{re}&{\mathcal{R}}^{e}&(12,34)=\widetilde{\mathcal{R}}^{0}(12,34)+\sum_{5678}^{\tau}\widetilde{\mathcal{R}}^{0}(12,56)\Big[\mathcal{U}^{e}(56,78)+\nonumber\\
&+&\widetilde{\mathcal{G}}^{-1}(7,5)\Sigma^{e}(6,8)+\Sigma^{e}(7,5)\widetilde{\mathcal{G}}^{-1}(6,8)\Big]{\mathcal{R}}^{e}(78,34).\nonumber\\
\end{eqnarray}
At $T=0$ it can be solved in the time blocking approximation \cite{Tselyaev1989,KamerdzhievTertychnyiTselyaev1997} which is obtained by introducing a time projection operator into the integral part of Eq. (\ref{re}). The analogous imaginary-time projection operator for the finite temperature case would look as follows:
\begin{eqnarray}
\Theta(12,34)&=&\delta_{\sigma_{1},-\sigma_{2}}\theta(\sigma_{1}\tau_{41})\theta(\sigma_{1}\tau_{32}).
\label{t-proj}
\end{eqnarray}
With such time projection, the triple Fourier transform onto the energy domain and integrating over two energy variables leads to the equation of the Dyson type with a single frequency at $T=0$. This allows for obtaining a clearly positive definite spectral (strength) functions and stable numerical schemes performing a sort of time ordering. However, it turns out that at $T>0$ this operator does not lead to a similar result. 

We found that, in order to reach the desired form of the kernel,  the imaginary-time projection operator has to be modified as follows: 
\begin{eqnarray}
\Theta(12,34;T)=\delta_{\sigma_{1},-\sigma_2}\theta(\sigma_{1}\tau_{41})\theta(\sigma_{1}\tau_{32})\times\nonumber\\
%&\times&\left\{n_{k_{2}}[\sigma_{k_{1}}(\varepsilon_{k_{2}}-\mu),T]\theta(\sigma_{k_{1}}\tau_{12})+\right.\nonumber\\
%&+&\left.n_{k_{1}}[\sigma_{k_{2}}(\varepsilon_{k_{1}}-\mu),T]\theta(\sigma_{k_{2}}\tau_{12})\right\},
\times[n(\sigma_{1}\varepsilon_{2},T)\theta(\sigma_{1}\tau_{12})
%\nonumber\\
+n(\sigma_{2}\varepsilon_{1},T)\theta(\sigma_{2}\tau_{12})],\nonumber\\
\label{t-proj-ft}
\end{eqnarray}
i.e. it needs to contain an additional multiplier with the dependence on the Fermi-Dirac occupation numbers, which turns to unity in the $T=0$ limit at the condition $\sigma_{1} =-\sigma_{2}$. Acting by the projection operator $\Theta(12,34;T)$ on the components of $\widetilde{\mathcal{R}}^{0}(12,34)$, we construct an operator $\widetilde{\mathcal{D}}$, which reads:
\begin{eqnarray}
\label{Operator D}
\widetilde{\mathcal{D}}(12,34) = \Theta(12,34;T)\widetilde{\mathcal{R}}^{0(\sigma_1\sigma_2)}(12,34) =
\nonumber\\
=\delta_{\sigma_1,-\sigma_2}\widetilde{\mathcal{G}}^{\sigma_1}(3,1)\widetilde{\mathcal{G}}^{\sigma_2}(2,4)\theta(\sigma_1\tau_{41})\theta(\sigma_1\tau_{32})\times\nonumber\\
\times[n(\sigma_1\varepsilon_2,T)\theta(\sigma_1\tau_{12})
%\nonumber\\
+
n(\sigma_2\varepsilon_1,T)\theta(\sigma_2\tau_{12})]\nonumber \\
\end{eqnarray}
and make a replacement $\widetilde{\mathcal{R}}^{0}\rightarrow \widetilde{\mathcal{D}}$
in the second term of Eq. (\ref{re}). In Eq. (\ref{Operator D}), $\sigma_{k}=+1(-1)$ for particle (hole) components. The multiplier $\delta_{\sigma_1,-\sigma_2}$ constrains the possible combinations of $(\sigma_1,\sigma_2)$ to be ($+1,-1$) and ($-1,+1$). A pair of states $\{12\}$ is regarded as a $ph$ ($hp$) pair if the energy difference $\varepsilon_{1}-\varepsilon_{2}$ is larger (smaller) than zero. As at $T=0$, the replacement of $\widetilde{\mathcal{R}}^{0}$ by $\widetilde{\mathcal{D}}$ corresponds to the elimination of configurations, which are more complex than $ph\otimes phonon$ ones. Thus, besides confining by only two-fermion correlation functions, in the leading approximation we keep the terms with only $ph$ and $ph\otimes phonon$ configurations and neglect higher-order terms.

After the triple Fourier transformation with respect to the imaginary-time,
\begin{eqnarray}
&{\mathcal{R}}^{e}_{12,34}&(\omega_{n},\varepsilon_{\ell},\varepsilon_{\ell'})=\frac{1}{8}\int_{-1/T}^{1/T}d\tau_{31}d\tau_{21}d\tau_{34}\times\nonumber\\
&&\times e^{i(\omega_{n}\tau_{31}+\varepsilon_{\ell}\tau_{21}+\varepsilon_{\ell'}\tau_{34})}{\mathcal{R}}^{e}(12,34),
\end{eqnarray}
the summation over the fermionic discrete variables $\ell$ and $\ell'$, 
\begin{eqnarray}
{\mathcal{R}}^{e}_{12,34}(\omega_{n})=T^{2}\sum_{\ell}\sum_{\ell'}{\mathcal{R}}^{e}_{12,34}(\omega_{n},\varepsilon_{\ell},\varepsilon_{\ell'}),\nonumber\\
\end{eqnarray}
and the analytical continuation to the real frequencies, we obtain 
\begin{equation}
\label{re1}{\mathcal{R}}^{e}_{12,34}(\omega)=\widetilde{\mathcal{R}}^{0}_{12,34}(\omega)
+\sum_{56,78}\widetilde{\mathcal{R}}^{0}_{12,56}(\omega){\Phi}_{56,78}(\omega){\mathcal{R}}^{e}_{78,34}(\omega),\nonumber\\
\end{equation}
where the uncorrelated propagator is given by:
\be
\label{Free Response}\widetilde{\mathcal{R}}^{0}_{12,34}(\omega)=\delta_{13}\delta_{24}\frac{n(\varepsilon_{2},T)-n(\varepsilon_{1},T)}{\omega-\varepsilon_{1}+\varepsilon_{2}}
\ee
and the particle-vibration coupling amplitude reads: 
\begin{eqnarray}
\label{phi}
{\Phi}_{12,34}(\omega)=\frac{\delta_{\sigma_{1},-\sigma_{2}}\sigma_{1}}{n(\varepsilon_{4},T)-n(\varepsilon_{3},T)}\times\nonumber\\
\times\sum_{56;\mu}\sum_{\eta_{\mu}=\pm 1}\eta_{\mu}\zeta^{\mu\eta_{\mu}}_{12,56}\zeta^{\mu\eta_{\mu}\ast}_{34,56}\times\nonumber\\
\times\frac{[N(\eta_{\mu}\Omega_{\mu})+n(\varepsilon_{6},T)]
[n(\varepsilon_{6}-\eta_{\mu}\Omega_{\mu},T)-n(\varepsilon_{5},T)]}{\omega-\varepsilon_{5}+\varepsilon_{6}-\eta_{\mu}\Omega_{\mu}}.\nonumber\\
\end{eqnarray}
Here we defined the phonon vertex matrices $\zeta^{\mu\eta_{\mu}}$ according to:
\begin{eqnarray}
\zeta^{\mu\eta_{\mu}}_{12,56}&=&\delta_{15}g^{\mu(\eta_{\mu})}_{62}-g^{\mu(\eta_{\mu})}_{15}\delta_{62}
%g^{m\eta_{m}}_{k_{1}k_{3}}&=&\delta_{\eta_{m},+1}g^{m}_{k_{1}k_{3}}+\delta_{\eta_{m},-1}g^{m\ast}_{k_{3}k_{1}}.
\end{eqnarray}
and the matrix elements of the phonon vertices $g^{\mu(\eta_{\mu})}$ \cite{Tselyaev1989,LitvinovaRingTselyaev2007} as:
\begin{equation}
g^{\mu(\sigma)}_{12}=\delta_{\sigma,+1}g^{\mu}_{12}+\delta_{\sigma,-1}g_{21}^{\mu\ast}.
\end{equation}
The phonon vertices $g_{12}^{\mu}$ are formally related to the transition densities $\rho_{12}^{\mu}$ 
\begin{equation}
g^{\mu}_{12}=\sum_{34}\widetilde{V}_{12,34}\rho^{\mu}_{34},
\end{equation}
which can be extracted from the full response function as described in Ref. \cite{WibowoLitvinova2018}, however, in practice it is sufficient to calculate them in the random phase approximation.
The static interaction, in the framework based on the covariant energy density functional (\ref{cedf}), obeys the well-known relation:
${\hat{\widetilde V}} = \delta^2 E[{\hat\rho,\phi_m}]/(\delta {\hat\rho} \delta {\hat\rho})$. In the case of effective mesons, whose masses and coupling vertices are adjusted to bulk properties of finite nuclei, the double counting of the particle-vibration coupling (PVC) is removed by correcting the PVC amplitude as follows: 
\begin{equation}
{\Phi}(\omega)\rightarrow\delta{\Phi}(\omega)={\Phi}(\omega)-{\Phi}(0),
\label{sub}
\end{equation}
i.e. by the subtraction of itself at $\omega=0$. This subtraction procedure and its role in the density functional based calculations is discussed in more detail in Refs. \cite{Tselyaev2013,LitvinovaTselyaev2007,LitvinovaRingTselyaev2007} for the case of $T=0$. At finite temperature we perform the subtraction in the complete analogy to the zero-temperature case.

Finally, the nuclear strength function $\tilde{S}(E)$ is defined as the response to a certain external probe associated with a one-body operator ${\hat V}^{0}$:
\begin{eqnarray}
\label{Strength}
\tilde{S}(E)&=&\frac{1}{1-e^{-E/T}}{S}(E),\nonumber \\
{S}(E)&=&-\lim_{\Delta\rightarrow+0}\frac{1}{\pi}\mbox{Im}\sum_{1234}V^{0\ast}_{12}\mathcal{R}_{12,34}(E+i\Delta)V^{0}_{34}.\nonumber\\
\end{eqnarray}
The exponential factor $\left[1-\exp(-E/T)\right]^{-1}$ appears due to the detailed balance between the absorption and emission parts of the strength function \cite{WibowoLitvinova2018,Sommermann1983,RingRobledoEgidoEtAl1984}. This factor brings a new feature to the finite-temperature strength function, for instance, it changes the low-energy behavior and the zero-energy limit of $\tilde{S}(E)$. In addition to the appearance of new poles in the response function due to the thermal unblocking, the exponential factor leads to a finite strength function at $E=0$, in contrast to the spectral density $S(E)$ which has zero limit at $E\to 0$. 

\section{Numerical details, results and discussion}
\label{sec:2}

In this work, we illustrate the performance of the developed approach named finite-temperature relativistic time-blocking approximation (FT-RTBA) for the response of the even-even spherical nuclei $^{48}$Ca and $^{100,120,132}$Sn. The calculation scheme consists of the following steps. First,  the closed set of the RMF equations is solved using the  NL3 parametrization \cite{Lalazissis1997a} of the non-linear sigma-model with the thermal fermionic occupancies (\ref{fdo}). The obtained temperature-dependent single-particle Dirac spinors and the corresponding single-nucleon energies form the basis for subsequent calculations. 
Second, the finite-temperature relativistic random phase approximation (FT-RRPA) equations, which are equivalent to Eq. (\ref{BSE}) without the time-dependent kernel $\cal W$, are solved to obtain the phonon vertices $g^{\mu}$ and their frequencies $\Omega_{\mu}$. 
The set of phonons, 
%with the $J_m^{\pi_m}$ = 2$^+$, 3$^-$, 4$^+$, 5$^-$, 6$^+$ and frequencies $\omega_m \leq$ 20 MeV
together with the RMF single-particle basis, forms the $1p1h\otimes$phonon configurations for the particle-phonon coupling amplitude ${\Phi}(\omega)$.
Then,  Eq. (\ref{re1}) is solved in the truncated configuration space which includes excitations below 25 MeV and the total response function is computed by solving Eq. (\ref{FullResp}) in either momentum or configuration space as described in \cite{LitvinovaRingTselyaev2007,LitvinovaRingTselyaev2008} for the $T=0$ case. Finally, the strength function 
%${S}(E)$ 
is computed according to Eq. (\ref{Strength}) with external fields of the isoscalar (L=0,2) and electromagnetic (L=1) character:
\bea
V^{0}_{00} &=&  \sum\limits_{i=1}^Ar_i^2 Y_{00}(\Omega_i)\nonumber \\
V^{0}_{1M} &=& \frac{eN}{A}\sum\limits_{i=1}^Z r_iY_{1M}(\Omega_i) - \frac{eZ}{A}\sum\limits_{i=1}^N r_iY_{1M}(\Omega_i) \nonumber \\
V^{0}_{2M} &=& e \sum\limits_{i=1}^Ar_i^2 Y_{2M}(\Omega_i).
\eea 

The particle-hole basis was limited by $\varepsilon_{ph}\leq100$ MeV and $\varepsilon_{\alpha h}\geq-1800$ MeV with respect to the positive-energy continuum. A direct verification with $\varepsilon_{ph}\leq300$ MeV eliminating the dipole spurious translational mode completely showed that the physical states of the excitation spectra converge reasonably well with $\varepsilon_{ph}\leq100$ MeV. The values of the smearing parameter $\Delta = 500$ keV and $\Delta = 200$ keV were adopted for the calculations of the strength functions for $^{48}$Ca and $^{100,120,132}$Sn, respectively. The phonon space was formed of the vibrations with quantum numbers of spin and parity $J^{\pi}=2^{+},\;3^{-},\;4^{+},\;5^{-},\;6^{+}$ below the energy cutoff, which amounts to 15 and 20 MeV for heavy and medium-mass nuclei, respectively. This cutoff is justified by our previous calculations at $T=0$ with the subtraction (\ref{sub}) which minimizes the the high-energy phonon contribution. An additional truncation was done according to the values of the reduced transition probabilities of the corresponding electromagnetic transitions. The modes with the values of the reduced transition probabilities $B(EL)$ less than 5\% of the maximal one (for each $J^{\pi}$) were neglected. We have put a condition that the same truncation criteria on the phonon energy, $J^{\pi}$ and the reduced transition probability should be kept for all temperature regimes in order to make a fair comparison of the calculated strength distributions. At zero and low temperatures varying these limits would almost not change the resulting strength distributions and, therefore, we typically adopt these cutoff values for $T=0$ calculations. At high temperatures, as a consequence of the significant thermal unblocking, we see the appearance of many additional phonon modes. This effect is illustrated below in Figs. \ref{fig:4}-\ref{fig:6} for $J^{\pi} = 0^{+},\;1^{-},\;2^{+}$ and it is similar for all $J^{\pi}$ values. On average, at $T\sim 5-6$ MeV the number of phonon modes in the truncated model space becomes by an order of magnitude larger than at $T= 0$.  We note that at high temperatures the saturation of the results with respect to the $B(EL)$ cutoff is rather slow, but the 5\% cutoff, which is at the limit of our current computational capabilities, is still very reasonable. We have checked that by varying the cutoff value between 10\% and 5\%. Another truncation was made on the absolute values of the numerator of Eq. (\ref{Free Response}), so that the contributions with $|n(\varepsilon_{2},T)-n(\varepsilon_{1},T)|\leq0.01$
were excluded from the BSE. FT-RTBA calculations at high temperatures are otherwise very prohibitive, while at low and moderate temperatures this truncation does not affect the results. We found, however, that in some cases at high temperature values this truncation can cause a noticeable contribution of the spurious translational mode in the lowest-energy parts of the dipole spectra, therefore, the truncation on the occupation numbers was released as much as possible to remove this contribution where it appeared.

\begin{figure}
\resizebox{0.47\textwidth}{!}{\includegraphics{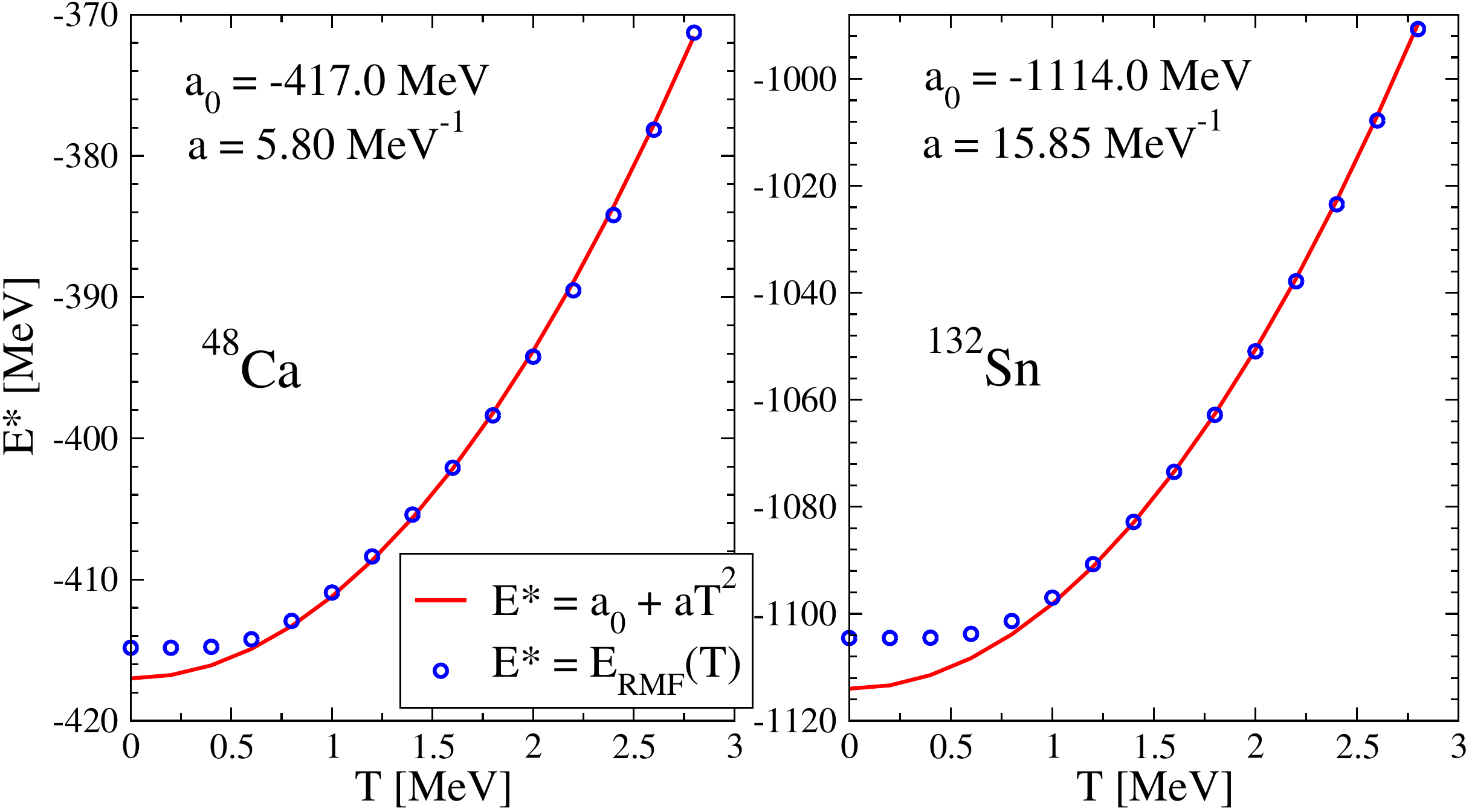}}
%\vspace{5cm}       % Give the correct figure height in cm
\caption{Energies of the thermally excited compound nuclei $^{48}$Ca and $^{132}$Sn as functions of temperature: RMF (blue circles) and parabolic fits (red curves).}
\label{fig:2}       % Give a unique label
\end{figure}

%%%%%%%%%%%%%
An illustration of the thermal RMF calculations for the compound nuclei $^{48}$Ca and $^{132}$Sn at $J^{\pi} = 0^+$ is given in Fig. \ref{fig:2}. As it follows from Eqs. (\ref{cedf},\ref{fdo}), the effect of finite temperature on the total (binding) energy of a thermally excited nucleus is mainly induced by the change of the fermionic occupation numbers. Their direct effect is caused by promoting the fermions to higher-energy orbits, which obviously increases the total energy. Another, indirect, contribution is related to the change of the nucleonic densities which play the role of sources for the meson and photon fields \cite{WibowoLitvinova2018}. The changes in the bosonic fields, in turn, translate to the nucleons via the self-consistent set of the thermal RMF equations. Thus, the thermodynamical equilibrium is achieved throughout the iteration procedure. As one can observe in Fig. \ref{fig:2}, the dependence $E^{\ast}(T)$ is nearly parabolic, as it is expected for the non-interacting Fermi gas, except for the lowest temperature values. 
The flat behavior of the excitation energy at these temperatures is a result of the presence of the large shell gaps right above the Fermi surface in the doubly-magic nuclei: small temperatures are insufficient to promote the nucleons over the shell gaps. At $T\geq$ 1 MeV the dependence  $E^{\ast}(T)$ is very well approximated by the parabolic fits, which are also shown in Fig. \ref{fig:2}. The corresponding coefficient at the quadratic term is found to be in a very good agreement with the empirical Fermi gas level density parameter $a = A/k$, where $8\leq k\leq 12$ \cite{Harakeh2001}.

\begin{figure}
\resizebox{0.47\textwidth}{!}{\includegraphics{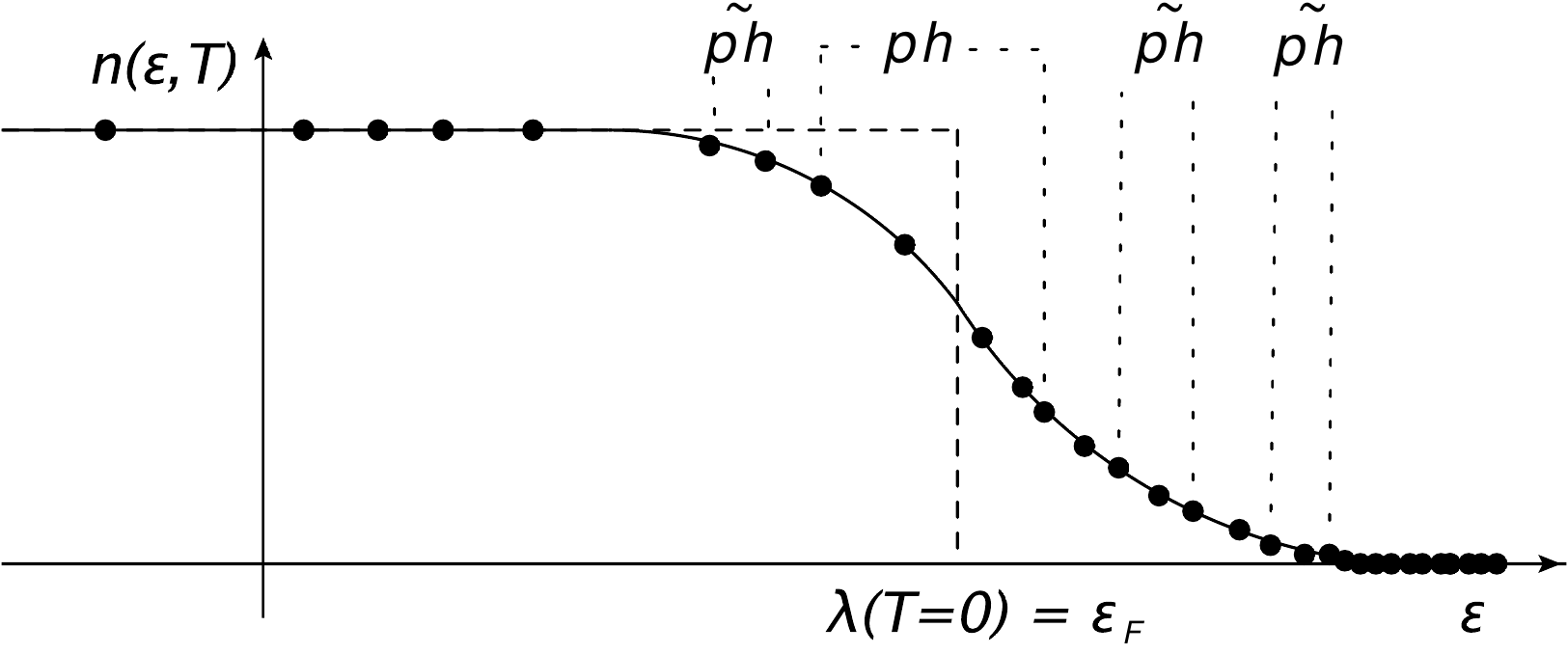}}
%\vspace{5cm}       % Give the correct figure height in cm
\caption{The appearance of thermally unblocked pairs with both single-nucleon states below or above the Fermi energy $\varepsilon_{F}$. Here $\widetilde{ph}$ stands for the particle-particle ($pp$) and hole-hole ($hh$) fermionic pairs with the non-vanishing uncorrelated propagator (\ref{Free Response}).}
\label{fig:3}       % Give a unique label
\end{figure}

Fig. \ref{fig:3} explains qualitatively the new mechanism of formation of nuclear excitations, which emerges with the temperature growth. Due to the thermal occupation factors of the single-nucleon orbits (\ref{fdo}), a sharp borderline between particle and hole states becomes diffuse as all states acquire a non-zero probability to be occupied. Therefore, the notion of particle-hole pairs adopts a conditional character. As far as the response is concerned, in the simplest approximation it is governed by the uncorrelated propagator of Eq. (\ref{Free Response}), which has formally the same ansatz as at $T=0$. At zero temperature, when the occupation factors take the values of zero above the Fermi energy and unity below the Fermi energy, the only non-vanishing contributions come, obviously, from the pairs of single-nucleon states located on different sides from the Fermi surface (particle-hole, or $ph$ pairs). At finite temperature, due to the fractional occupancies, the uncorrelated propagator of Eq. (\ref{Free Response}) acquires contributions from all pairs of states, except for those of the identical states. This fact allows for an extended notion of the particle-hole pairs. As mentioned in the previous section, we  regard a pair of states $\{12\}$ as a 'thermal' $ph$ ($hp$) pair if the energy difference $\varepsilon_{1}-\varepsilon_{2}$ is larger (smaller) than zero. Remarkably, in this context a single-particle state alone does not carry a particle or a hole character, but adopts it only being paired with another state. In Fig. \ref{fig:3} we have denoted as $ph$ the pairs which have a particle-hole character already at $T=0$ and as $\widetilde{ph}$ the new, or 'thermal' $ph$-pairs which are classified as particle-particle or hole-hole ones at $T=0$ and do not contribute to the zero-temperature uncorrelated propagator and, thus, to the zero-temperature response. 
\begin{figure}
\resizebox{0.47\textwidth}{!}{\includegraphics{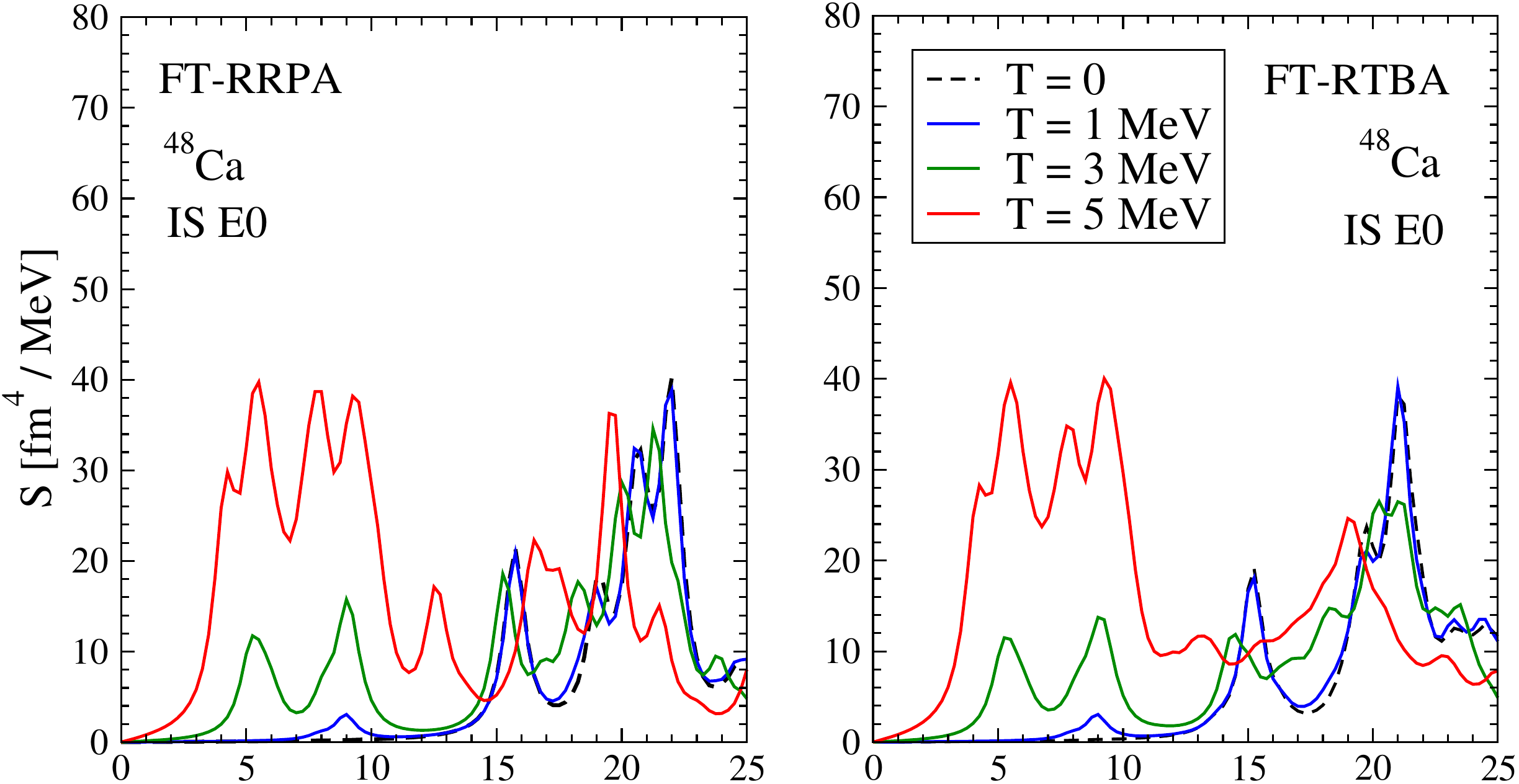}}
\resizebox{0.47\textwidth}{!}{\includegraphics{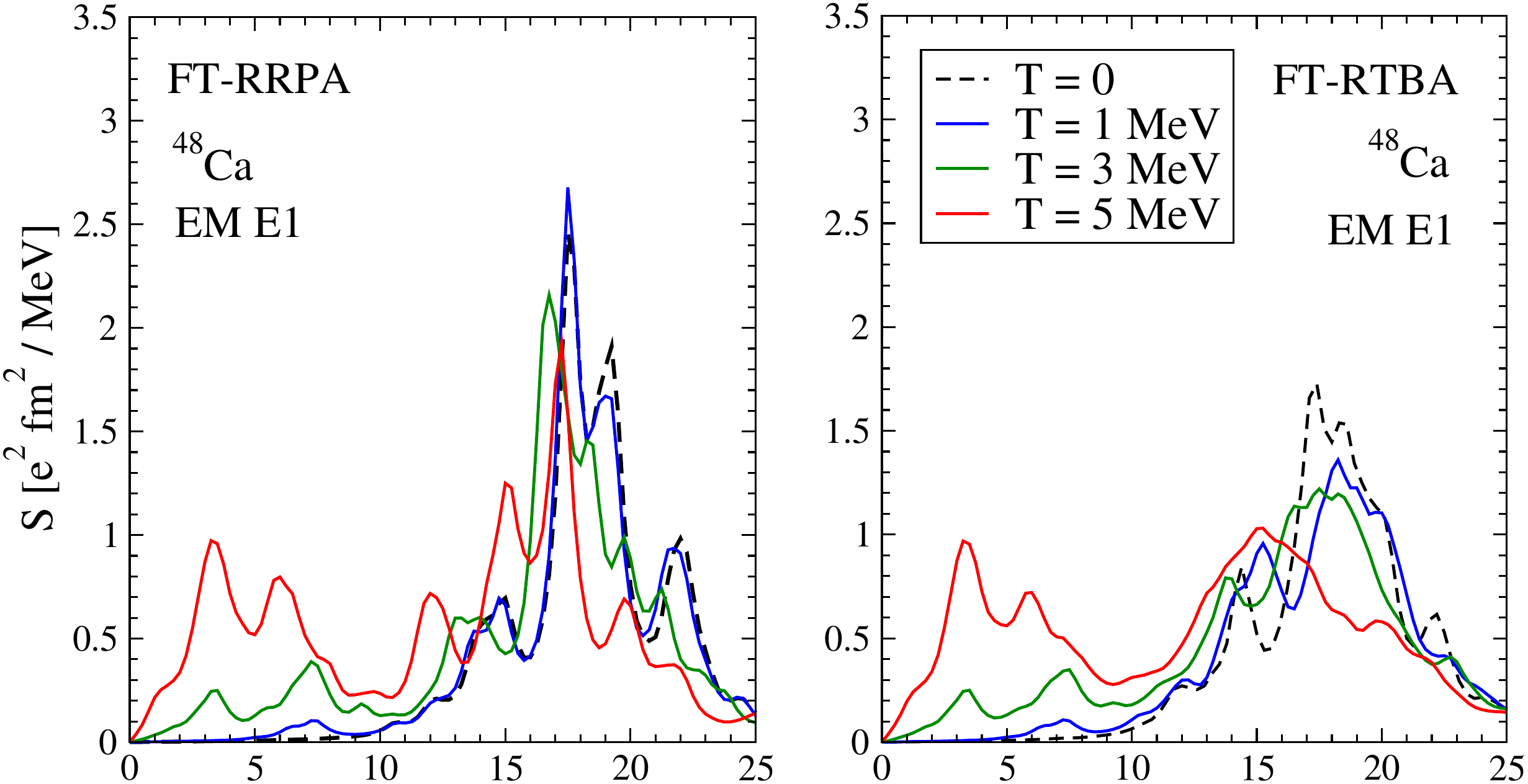}}
\resizebox{0.47\textwidth}{!}{\includegraphics{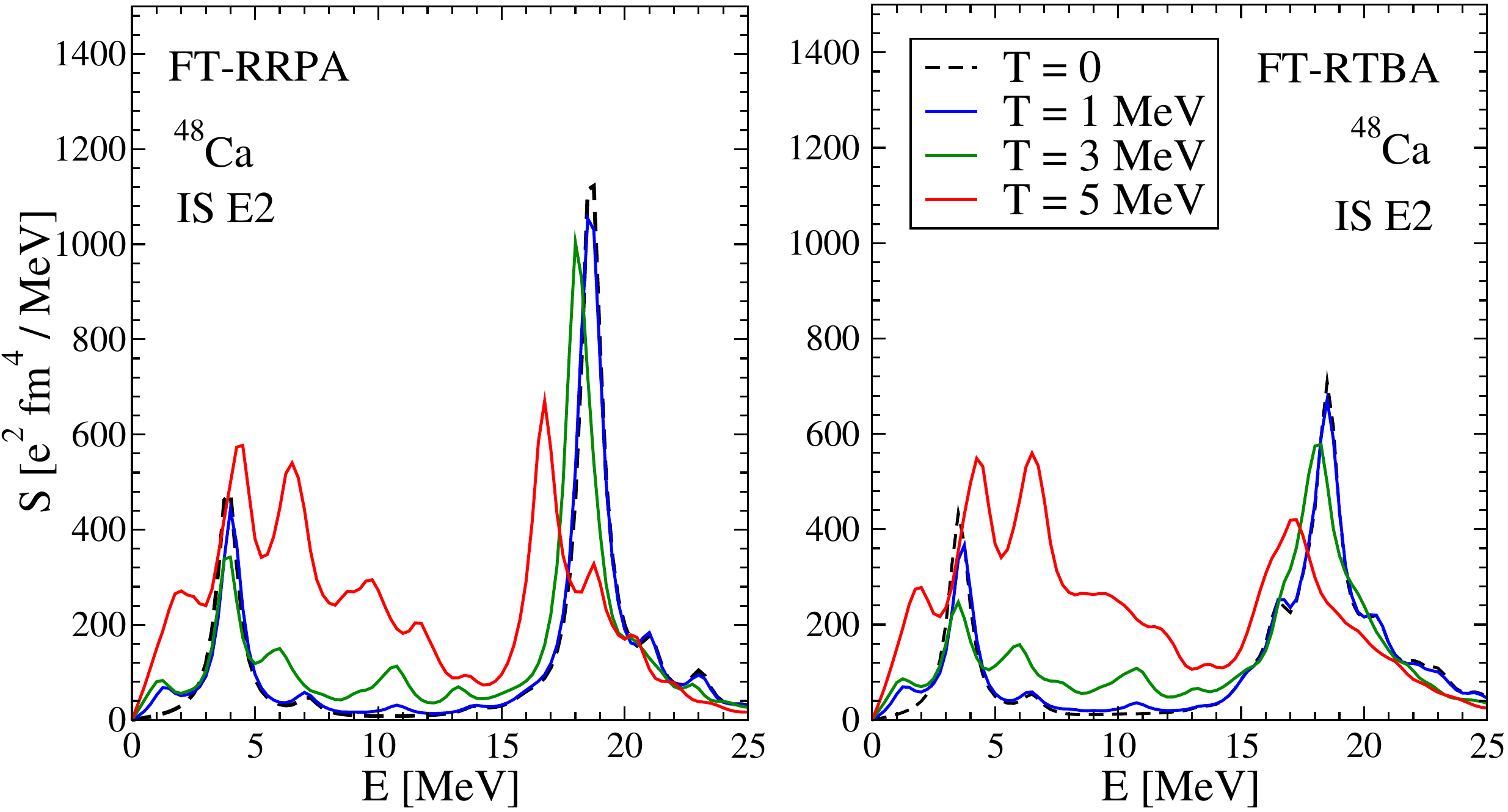}}
%\vspace{5cm}       % Give the correct figure height in cm
\caption{Temperature dependence of the isoscalar monopole IS E0 (top), electromagnetic dipole EM E1 (middle) and isoscalar quadrupole IS E2 (bottom) strength distributions in $^{48}$Ca. Left panels show FT-RRPA and right panels FT-RTBA results.}
\label{fig:4}       % Give a unique label
\end{figure}

The isoscalar monopole, electromagnetic dipole, and isoscalar quadrupole strength distributions were computed for the doubly-magic $^{48}$Ca in a wide range of temperatures, according to Eqs. (\ref{re1} - \ref{Strength},\ref{FullResp}) with the value of the imaginary part of the energy variable (smearing parameter) $\Delta = 500$ keV. The results of the FT-RTBA calculations  are displayed in the right panels of Fig. \ref{fig:4} and can be compared to the FT-RRPA results given in the left panels. At $T=0$ RTBA provides a very reasonable description of the strength distributions in $^{48}$Ca, in agreement with data, see, for instance, a direct comparison for the dipole response in our recent work \cite{EgorovaLitvinova2016}. This can serve as a good benchmark of the theory and, together with the thermal mean-field calculations presented above, justify the reliability of our approach to the strength distributions at finite temperature.
The isoscalar monopole response (ISMR) shown in the top panels of Fig. \ref{fig:4} represents the so-called breathing mode, or the compression mode, and serves as one of the most important nuclear characteristics. It is directly related to the nuclear compressibility which is studied very extensively, by both theory and experiment. One of the main motivations for these studies is that the characteristics of response of finite nuclei to compression can be translated or extrapolated to the compressibility of the uniform nuclear matter. Its incompressibility parameter, $K_{\infty}$, is one of the most important parameters of the nuclear matter equation of state (EOS) and, thus, plays the key role in modeling astrophysical objects, from core collapse supernovae to neutron star mergers. The incompressibility of a finite nucleus with the mass number $A$, $K_A$, can be determined from the centroid of ISMR and related to the $K_{\infty}$ through the well-known leptodermous expansion. The possibility of such extraction of $K_{\infty}$ uniquely is, however, very much debated as it leads to different results if different nuclei are used for this procedure. In particular, nuclear superfluidity seems to play a role in nuclear compressibility, because the results of analyses based on the monopole responses of closed-shell and open-shell nuclei vary, see, for instance, a recent review \cite{Garg2018}. As far as the astrophysical aspect of nuclear compressibility is concerned, its temperature dependence could be, in principle, another factor to be taken into account.  One can see, comparing the two top panels of Fig. \ref{fig:4}, that PVC effects are not strongly manifested in the case of monopole response at $T=0$. This is a well-known phenomenon related to a significant cancellations of terms with the self-energy and induced interaction in Eq. (\ref{phi}) in the monopole channel \cite{LitvinovaRingTselyaev2007}.  At $T=1$ MeV both FT-RRPA and FT-RTBA strength distributions demonstrate the appearance of a new soft mode at $E\approx$ 9 MeV. This is already a clear manifestation of the thermal unblocking mechanism discussed above. At $T=3$ MeV we observe first drastic changes in the monopole strength distribution: the soft mode at 9 MeV strengthens considerably while a new prominent mode emerges at  $E\approx$ 5 MeV. The locations of these states reveal the presence of  $0\hbar\omega$ excitations which are barely possible at $T=0$. Another new effect at this temperature is enforcing the fragmentation of the main high-energy peak whose origin stems from $2\hbar\omega$ excitations. The latter two features -- the emergence of strong low-lying states and the reinforcement of fragmentation of the high-energy peak - become even more amplified at $T=5$ MeV, moreover, the low-energy strength begins to dominate. In addition, the entire strength distribution shifts toward lower energies. Such changes in the isoscalar monopole strength distribution obviously affect its centroid, that, in turn, means a  change of the nuclear compressibility. We do not go into more details of the nuclear compressibility here, however, point out that its sensitivity to the temperature increase might be studied elsewhere more extensively for the temperature regimes which are relevant for astrophysical processes.

It can be seen from the middle panels of Fig. \ref{fig:4} that the main thermal effects which are observed for the case of ISMR, are also visible in the dipole channel. At zero and low temperatures, the latter is dominated by the giant dipole resonance (GDR), which is formed by collective oscillations of protons and neutrons against each other and centered at about $E_{GDR}=18A^{-1/3}+25A^{-1/6}$ MeV \cite{Bortignon1998}.
In the context of its temperature dependence, the dipole response was studied most extensively in experiments with heavy ion collisions \cite{Gaardhoje1984a,GaardhojeEllegaardHerskindEtAl1986,Bracco1989,Ramakrishnan1996,Mattiuzzi1997,Heckman2003,Santonocito2006,Wieland2006,Bracco2007}. The general conclusions from these studies are (i) a fast growth of the GDR's width with temperature and (ii) disappearance of the high-frequency collective oscillation at high temperatures. 

As discussed in detail in Ref. \cite{Bortignon1998}, the latter phenomenon occurs largely due to the particle emission which is not taken into account consistently in our approach and should be included with the exact continuum as it is done, for instance, in Ref. \cite{LitvinovaBelov2013}. However, even in the present calculations we observe a reduction of the coherence of the GDR with temperature.
 Indeed, one can see that, in the case of $^{48}$Ca, at $T=5$ MeV the soft mode starts to dominate over the high-frequency mode while the latter becomes more fragmented. At higher temperatures these effects become even more pronounced, see, for instance, an example of our calculations in Refs. \cite{LitvinovaWibowo2018,WibowoLitvinova2018}. 

Another enhancement of the low-energy strength  comes with the exponential factor in Eq. (\ref{Strength}) which is analyzed below. Experimental and theoretical studies of the nuclear dipole response also have far reaching broader impacts. The electric dipole polarizability, which is proportional to the $m_{-1}$ moment of the dipole strength distribution, provides another very important aspect of the nuclear EOS, namely the slope of its symmetry energy, see, for instance, a recent review \cite{Roca-Maza2018}. As the  $m_{-1}$ moment of the dipole strength distribution is particularly sensitive to the low-energy fraction of the strength, it should be also sensitive to the temperature increase, which can become crucial for the symmetry-energy sector of the EOS at high-temperature regimes.

The specific feature of a typical isoscalar quadrupole response (ISQR) at $T=0$ is the presence of two main collective structures: one at low (few MeV) and one at high energy. The properties of the ISQR are also of a broader interest as it is sensitive to the nuclear matter incompressibility, provides a constraint on the effective mass, and contributes to determining the nuclear symmetry energy and the neutron skin thickness \cite{Roca-Maza2018}. 
 In the context of theoretical models for nuclear response beyond the random phase approximation, the low-lying quadrupole mode is of a particular interest, because, together with its octupole counterpart, it provides the most important contribution to the PVC mechanism. The low-energy collective quadrupole and octupole phonons couple most strongly to the single-particle degrees of freedom and produce the largest fraction of the fragmentation 
of both single-particle and collective nuclear states. The ISQR in $^{48}$Ca at various temperatures is displayed in the bottom panels of Fig. \ref{fig:4}, where FT-RTBA results can be compared to those of FT-RRPA. The first thermal unblocking effect can be seen already at $T=1$ MeV as the formation of the new low-energy mode. With the temperature increase, i.e. at $0\leq T \leq3$ MeV the fragmentation of the high-energy quadrupole peak enforces gradually while at $T=5$ MeV it becomes visibly stronger. The latter is common for all $0^+, 1^-$ and $2^+$ strength distributions and related to the tremendously enhanced low-energy strength of all multipolarities $J^{\pi}=2^{+},\;3^{-},\;4^{+},\;5^{-},\;6^{+}$ included into the PVC amplitude (\ref{phi}).  These new strong low-lying modes tend to also couple strongly to the single-particle degrees of freedom and to contribute significantly to the $\Phi(\omega)$ amplitude, while the quadrupole contribution is one of the dominant ones. 

\begin{figure}
\resizebox{0.47\textwidth}{!}{\includegraphics{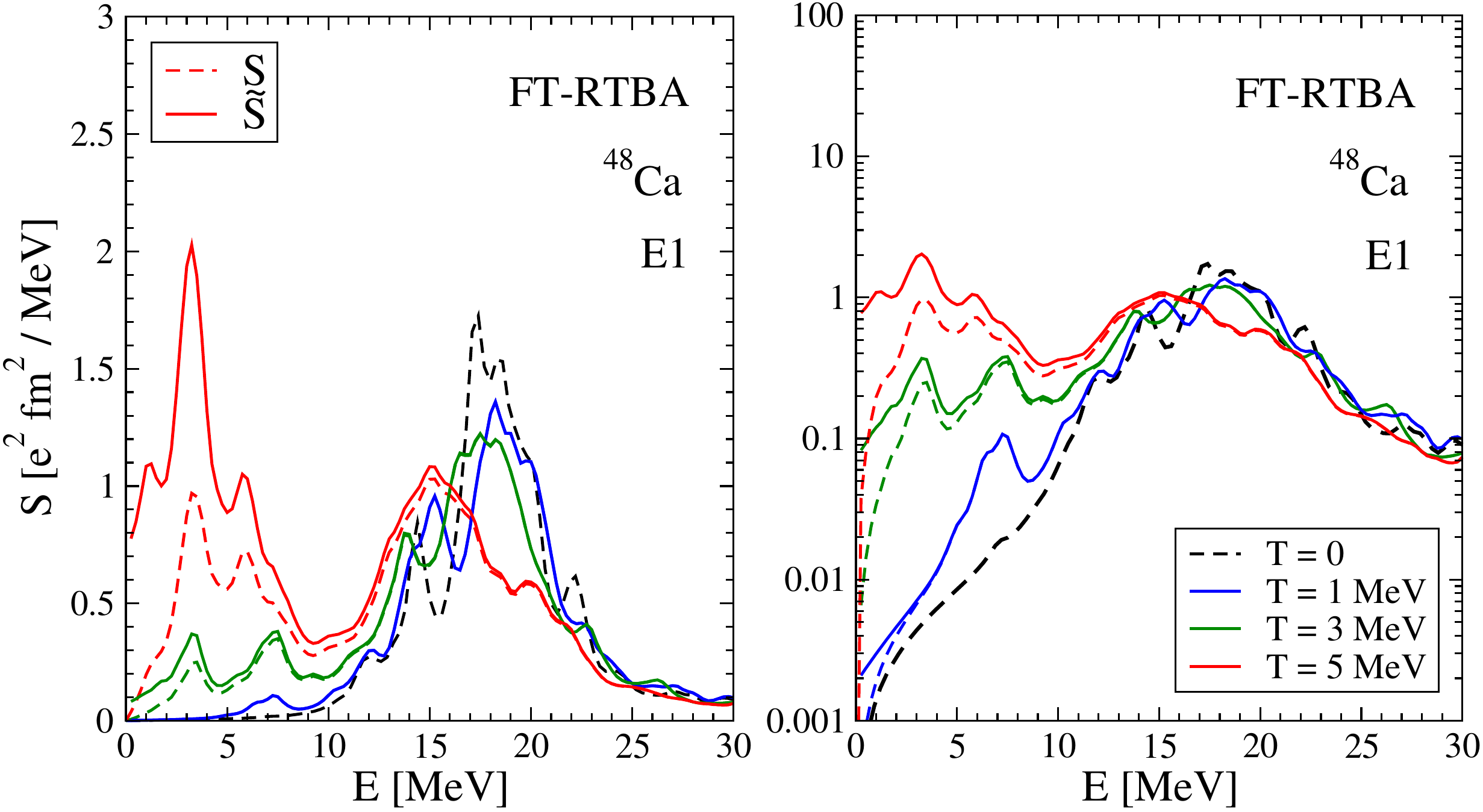}}
%\vspace{5cm}       % Give the correct figure height in cm
\caption{The role of the exponential factor in the finite-temperature strength distribution: an example of the FT-RTBA dipole strength in $^{48}$Ca.}
\label{fig:5}       % Give a unique label
\end{figure}

Here it should be noted that in Fig. \ref{fig:4} we plot not the final strength functions, but their zero-temperature analogs $S(E)$ of Eq. (\ref{Strength}) called spectral densities.  The strength function at finite temperature ${\tilde S}(E)$ has to obey the detailed balance between absorption and emission and, thus, to be corrected by the exponential factor $\left[1-\exp(-E/T)\right]^{-1}$ which is singular at $E=0$. In order to illustrate the role of this new factor, Fig. \ref{fig:5} shows the FT-RTBA dipole strength distributions in $^{48}$Ca with and without this factor. For a better assessment of its role at low temperatures, the right panel displays the same strength functions as shown in the left panel, but on the logarithmic scale. One can notice that the high-frequency giant dipole resonance remains almost unaffected at all temperatures. The situation is, however, different for the low-energy part.
It can be seen that already at $T=1$ MeV the strength function acquires a finite value at $E=0$ because in the present case $S(0) =0$  (no physical poles appear at zero energy) and the influence of the exponential factor on the low-energy strength grows with temperature. At $T=5$ MeV, for instance, the low-energy peak receives an enhancement of a factor of two. Thus, although such high temperatures are not relevant for typical astrophysical situations, temperatures of $\sim1-2$ MeV can be reached in many cases, so that the exponential factor should not be neglected in the finite-temperature astrophysical modeling of quantities which are sensitive to the zero-energy limit of the strength functions. 

In this work we mostly focus on the spectral density $S(E)$ because it has a well-defined zero-energy limit. Self-consistent 
calculations for the dipole strength should demonstrate, in particular, the absence of the admixture of the spurious translational mode. The presence of the singular factor in the full strength function ${\tilde S}(E)$ which makes it finite at $E=0$ conceals the low-energy behavior of the distribution $S(E)$ and makes an assessment of the quality of the numerical implementation difficult. We note that the exponential factor gives a very small contribution at temperatures below 4 MeV and can be easily included for a comparison with experimental data when they are available for the low-lying strength. By these reasons we found showing $S(E)$ more informative than the full strength function and illustrated the influence of the exponential factor in Fig. 5 as an example of its generic behavior.
For a reliable prediction of the low-energy strength distributions at finite temperatures at least one more aspect, namely the (single-particle) continuum should be included into the description, because it can further modify the trend of the low-energy behavior of the strength distributions of all multipoles \cite{LitvinovaBelov2013}. 
\begin{figure}
%\vspace*{5cm}       % Give the correct figure height in cm
\resizebox{0.47\textwidth}{!}{\includegraphics{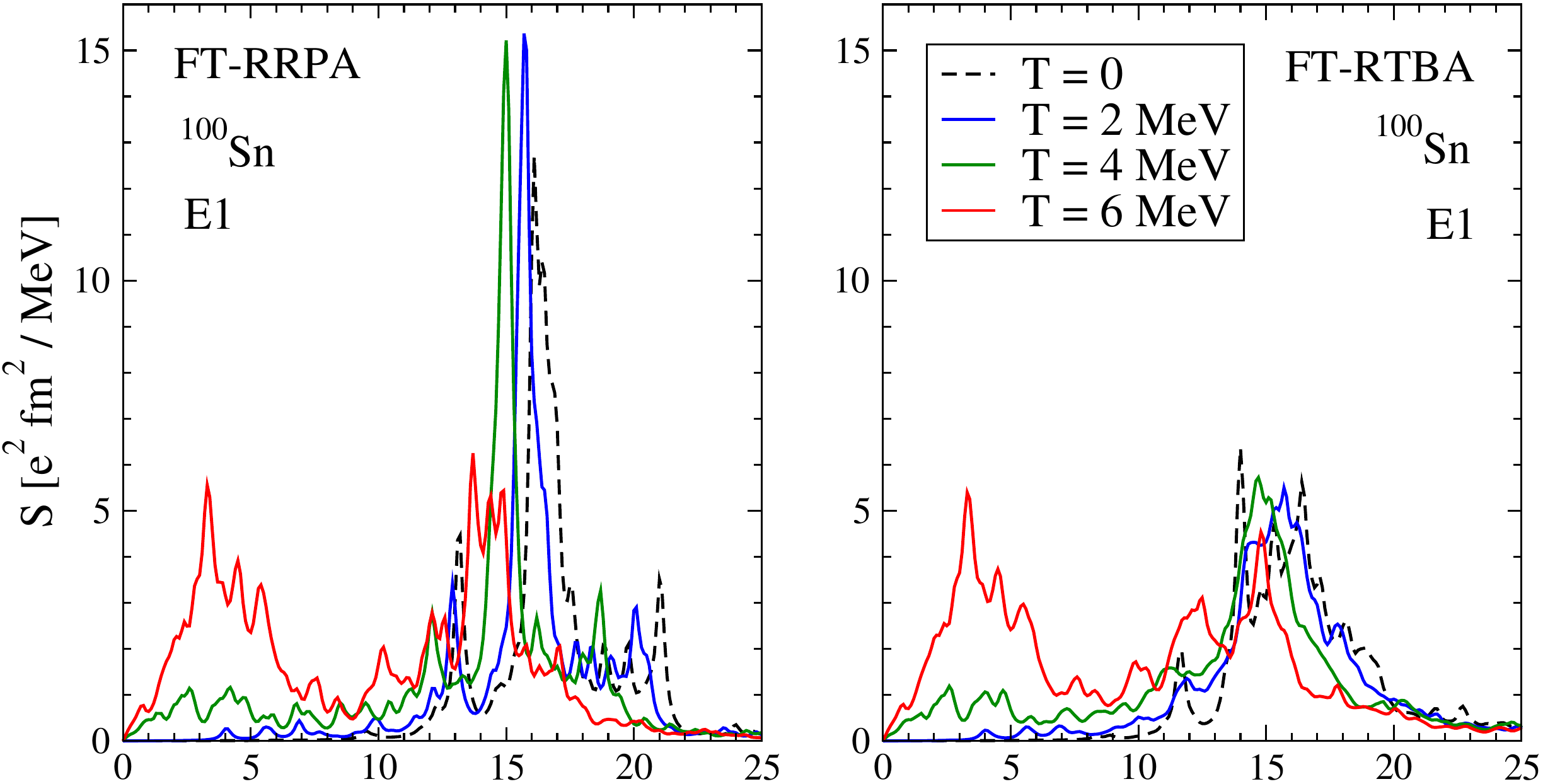}}
\resizebox{0.47\textwidth}{!}{\includegraphics{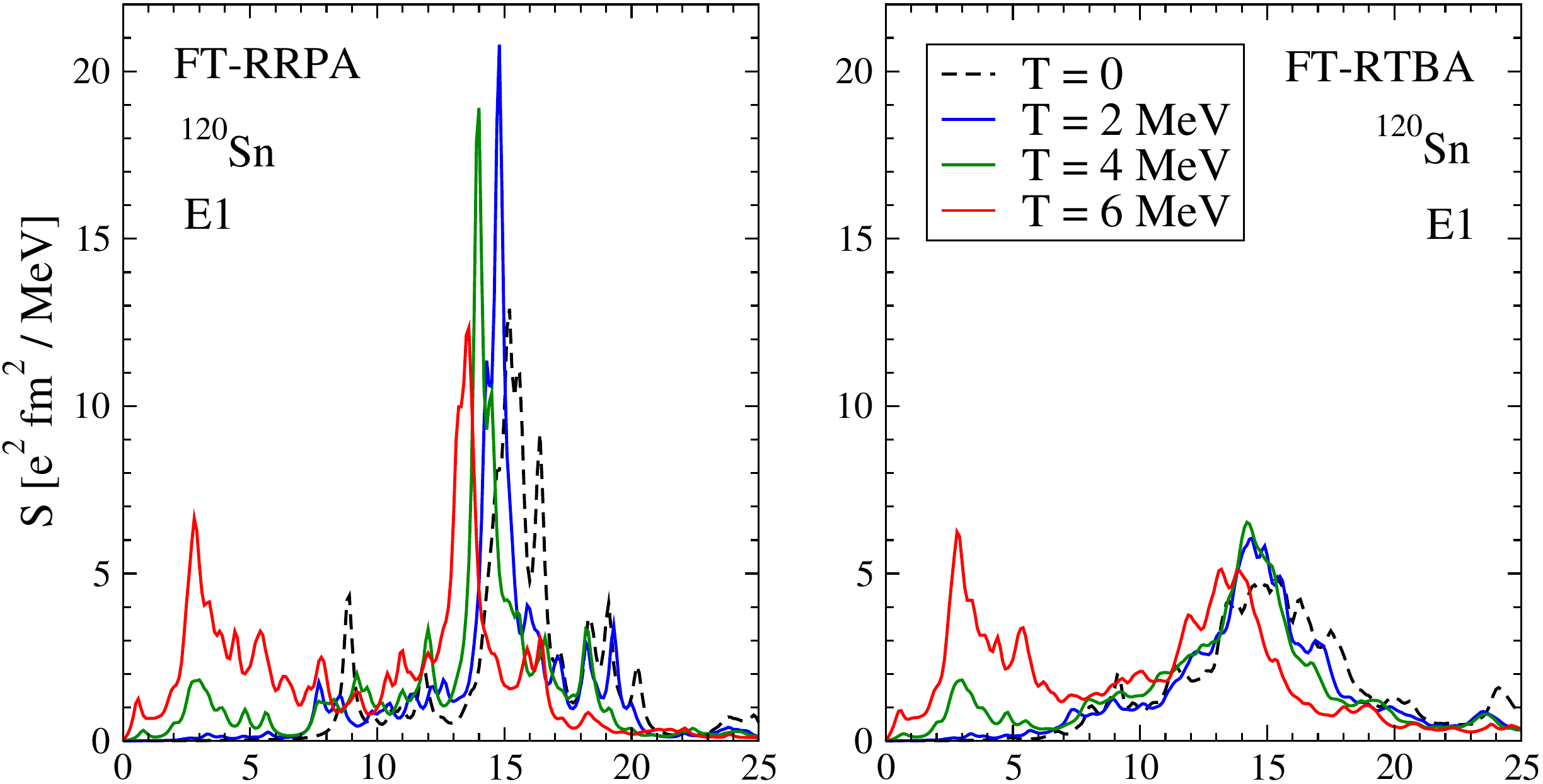}}
\resizebox{0.47\textwidth}{!}{\includegraphics{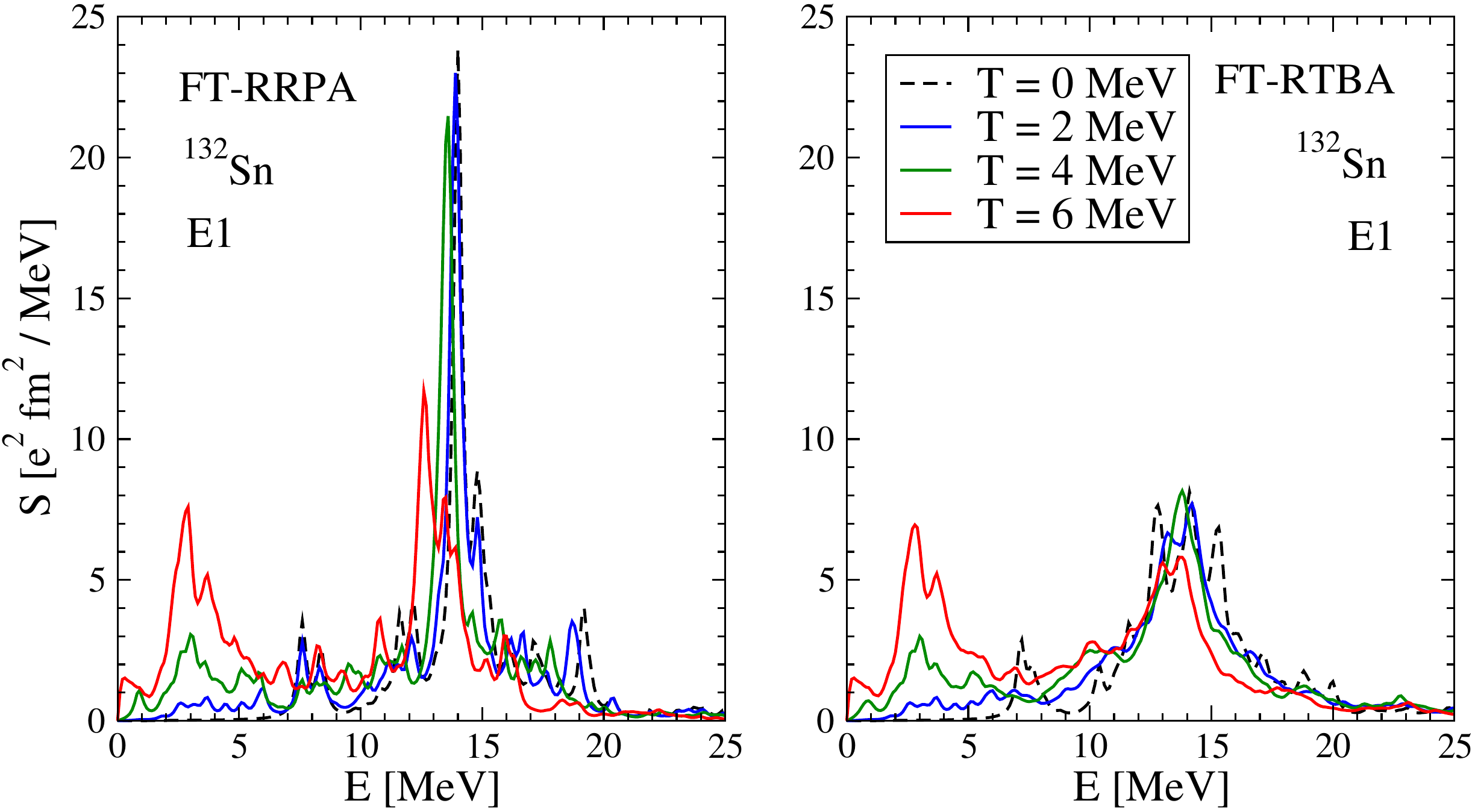}}
\caption{The giant dipole resonance in $^{100,120,132}$Sn (top, middle and bottom, respectively) as a function of temperature. Right panels display FT-RTBA calculations, to be compared to the left panels with FT-RRPA ones.}
\label{fig:6}       % Give a unique label
\end{figure}

The picture of the temperature evolution of the nuclear response would not be complete without considering heavier nuclear systems. Therefore, we have performed calculations for some isotopes of the tin chain. The nucleus $^{120}$Sn is of special interest because of the availability of experimental data for its giant dipole resonance in terms of the temperature-dependent width \cite{Fultz1969,Heckman2003,Ramakrishnan1996}. 
We have also considered the neutron-rich doubly magic $^{132}$Sn because of its particular importance for the astrophysical r-process nucleosynthesis, so that the precise knowledge about its strength functions is required for the r-process modeling \cite{MumpowerSurmanMcLaughlinEtAl2016}. The neutron-deficient $^{100}$Sn was chosen in order to find out whether the temperature evolution of the nuclear response is different in the neutron-rich and neutron-deficient cases. The results of our calculations for the dipole response of these three nuclei with the smearing parameter $\Delta = 200$ keV are displayed in Fig. \ref{fig:6}. One can observe immediately that the doubly-magic $^{100,132}$Sn isotopes reveal a trend in the temperature evolution of the dipole strength, which is similar to one already seen in $^{48}$Ca. The situation with the open-shell $^{120}$Sn is somewhat different in the low-temperature regime. This nucleus is superfluid at $T=0$, and the superfluidity effects typically produce a more spread strength distribution because of the presence of more poles in the response function. However, above the critical temperature of $T_c\approx$ 0.66 MeV the superfluidity, in its traditional understanding within the Bardeen-Cooper-Schrieffer or Hartree-(Fock)-Bogoliubov models, vanishes, while at $T=1$ MeV the thermal unblocking is not yet well pronounced. As a result, the dipole strength in $^{120}$Sn at $T=1$ MeV is less broad than at $T=0$, in contrast to the case of the closed-shell nuclei. Otherwise, at $T\geq$ 1 MeV the strength distributions in all three nuclei evolve in a similar manner. As in the case of 
$^{48}$Ca, we observe a gradual increase of the fragmentation and broadening of the high-energy peak, due to both Landau damping and PVC,  and the formation of the enhanced strength at low energies. Comparing the neutron-balanced $^{120}$Sn, the neutron-rich $^{132}$Sn and the neutron-deficient $^{100}$Sn nuclei, one can see that the proton-neutron composition does not play the major role in the temperature evolution of the dipole strength. In particular, the formation of the new pronounced low-energy structure at high temperature is very similar for the three considered tin isotopes. This means that this structure is not associated with the pygmy resonance as a neutron skin oscillation, but has a purely thermal unblocking origin. Indeed, from the Fig. \ref{fig:3} it is clear that in the calculations within a sufficiently large single-particle basis the number of $\widetilde{ph}$ pairs above the Fermi surface can grow tremendously with the temperature increase and form a coherent low-lying mode, because many of such pairs can have close values of the energy differences which enter the denominator of Eq. (\ref{Free Response}). Accounting for the single-particle continuum is expected to provide a more accurate description of the low-energy strength and will be addressed in future work. Another interesting aspect of this new low-energy structure is a relatively weak fragmentation due to the PVC mechanism, which is obtained for the strength functions of all multipolarities. As it was discussed in Refs. \cite{LitvinovaWibowo2018,WibowoLitvinova2018}, this can possibly occur due to missing contributions of the PVC-induced ground state correlations \cite{KamerdzhievTertychnyiTselyaev1997}, which should be considered among other future directions.

%%%%%% Gamma and EWSR: modified 

The width of the strength distribution and the energy weighted sum rule (EWSR) are the most important integral characteristics of the GDR which are often addressed in theoretical and experimental studies. For example, they help constraining theoretical approaches because of their almost model-independent character. In theoretical calculations, verification of the sum rules usually helps in testing the consistency of numerical implementations. 

\begin{figure}
\resizebox{0.48\textwidth}{!}{\includegraphics{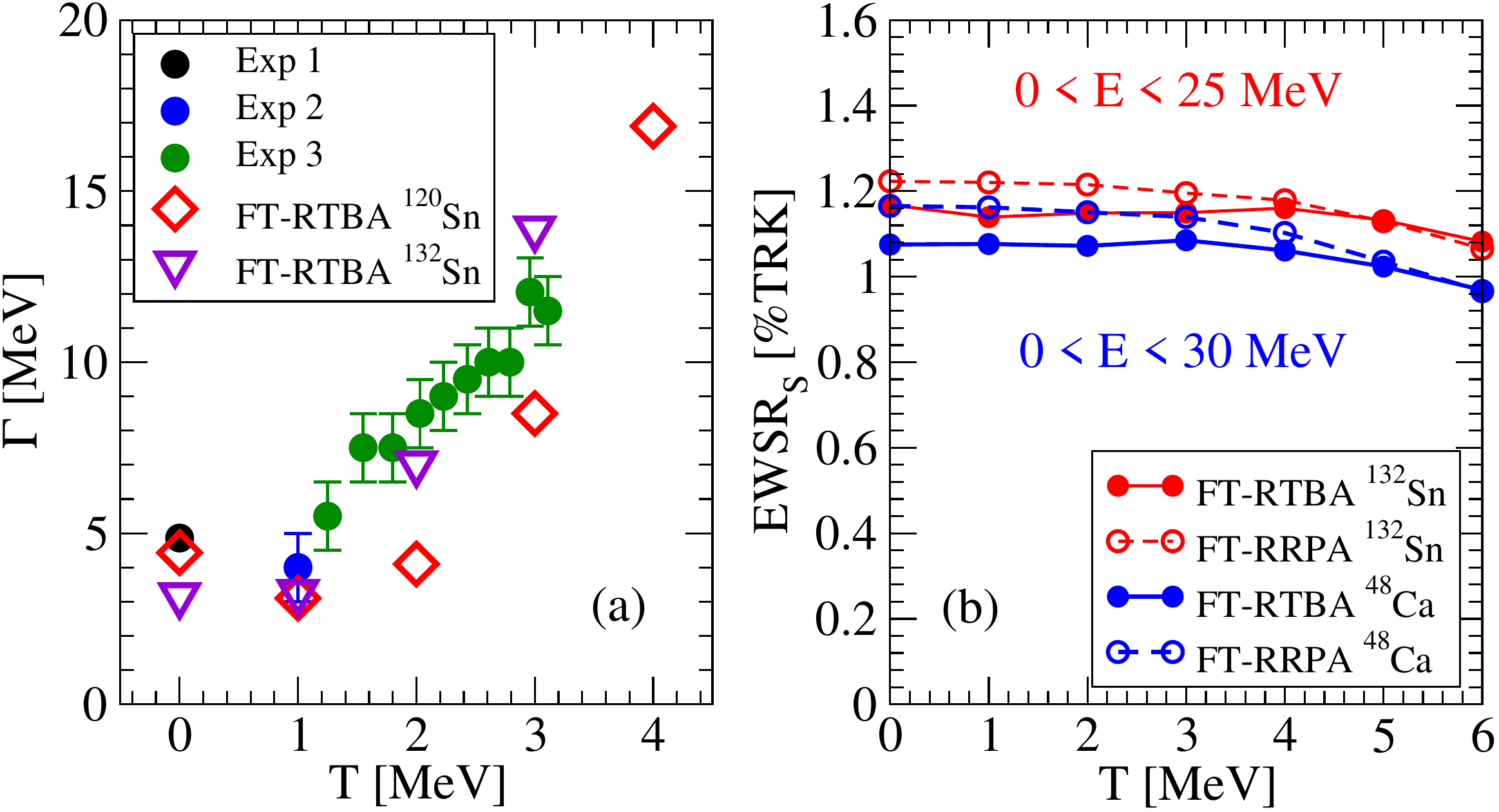}}
\caption{Integral characteristics of the giant dipole resonance: the GDR's width for $^{120,132}$Sn as a function of temperature (left panel), in comparison with experimental data \cite{Fultz1969,Heckman2003,Ramakrishnan1996} for $^{120}$Sn, and the energy-weighted sum rule (EWSR) in the percentage with respect to the TRK sum rule for $^{48}$Ca and $^{132}$Sn at various temperatures up to $T=6$ MeV (right panel). The figure is adopted from Ref. \cite{LitvinovaWibowo2018}.}
\label{fig:7}
\end{figure}

In order to illustrate the present case, the evolution of GDR's width $\Gamma(T)$ with  temperature obtained in FT-RTBA for $^{120}$Sn and $^{132}$Sn nuclei is shown together with experimental data in the left panel of Fig. \ref{fig:7}. Experimental data  which are available only for $^{120}$Sn are given as well. The theoretical values for the widths at $T=0$ are provided by our earlier calculations \cite{LitvinovaRingTselyaev2008,LitvinovaRingTselyaev2007}, respectively. %even at T=0. 
As follows from Fig. \ref{fig:2}, the thermal unblocking effects hardly appear at $T\leq1$ MeV in $^{132}$Sn
%and the reason can be understood by looking at the 
because of its specific shell structure, namely, the presence of the large shell gap in the vicinity of the Fermi energy in both proton and neutron subsystems. In particular, for protons, which form the $Z=50$ closed shell and have the next available orbitals only in the next major shell, temperatures below 1 MeV  are not sufficient to promote particles over the shell gap with a sizable occupancy. For neutrons in $^{132}$Sn as well as for protons in $^{120}$Sn the situation is similar. The lowest orbit available for neutrons in $^{120}$Sn is the intruder $1h_{11/2}$ state where particles get promoted relatively easily, but the next shell gap occurs right after this orbit. As a result, at $T=1$ MeV there are still not many options for the $\widetilde{ph}$ pair formation and for the thermal unblocking. Together with the disappearance of superfluidity above the critical temperature $T_c\approx 0.66$ MeV, this explains, for instance, the unexpectedly small GDR's width at $T=1$ MeV in $^{120}$Sn reported in Ref. \cite{Heckman2003}. After $T=1$ MeV in $^{132}$Sn and $T=2$ MeV in $^{120}$Sn we observe a quick growth of $\Gamma(T)$ due to the formation of the low-energy shoulder of the dipole strength distribution and due to the reinforced fragmentation of the high-energy peak emerging from the finite-temperature effects in the PVC amplitude $\Phi(\omega)$.
In general, at all temperatures $T\geq$ 1 MeV  the low-energy shoulder of $^{132}$Sn is stronger, which leads to a larger overall width in $^{132}$Sn, compared to $^{120}$Sn. This occurs because of the more neutron-rich character of $^{132}$Sn. Notice here that, since our standard Lorentzian fit of the microscopic strength distribution fails in recognizing the distribution as a single peak structure at high temperatures, the GDR's widths for T$>$3 MeV in $^{132}$Sn and for T$>$4 MeV in $^{120}$Sn are not presented in Fig. \ref{fig:7}.  

%%%%%%%%%%%%
%Overall, the agreement on the GDR's width in $^{120}$Sn between the FT-RTBA calculations and data is very reasonable. 
%The only temperature regime which looks unexplained is around $T=2$ MeV, where high angular momenta and deformation effects become important \cite{Santonocito2006}. As %mentioned above, the possibility of having a non-zero spin for the initial compound nucleus is not included in the present calculations and should be considered in future work. 
At $T=0, T=1$ MeV and $T=3$ MeV the agreement on the GDR's width in $^{120}$Sn between the FT-RTBA calculations and data is very reasonable while at $T=2$ MeV FT-RTBA underestimates the experimental value of the width by a factor of two. 
Here we should note that effects of thermal shape fluctuations are not included into the present approach. In order to do this in a microscopic way, we would need to generalize our method to the deformed case and to consider a superposition of shapes and orientations, as discussed, for instance, in Ref. \cite{Bortignon1998}. As a very large number of shapes and orientations would be required, the authors of Refs. \cite{OrmandBortignonBrogliaEtAl1990,OrmandBortignonBroglia1996}, for instance, did that in an effective way using phenomenological relations for the GDR. However, as it is shown, for instance, in Refs. \cite{Bonche.1984,Lisboa.2016,Zhang.2017}, nuclear deformation tends to disappear at some critical temperature.  These studies agree on the upper bound of the critical temperature $T_c \approx 2.0-4.0$ MeV. This means that above those temperatures medium-mass and heavy nuclei reveal the tendency to take the spherical shape. On the other hand, one can see from Refs. \cite{OrmandBortignonBrogliaEtAl1990,OrmandBortignonBroglia1996} that, although the theory of thermal shape fluctuations describes the GDR's width at temperatures $1.5\leq T \leq 3$ MeV well, it does not reproduce  $\Gamma(T)$ in the low temperature regime at $T<1.5$ MeV in $^{120}$Sn. Our microscopic theory, however, shows a reasonable  agreement with data for $T=0$ and $T=1$ MeV. From these arguments we can conclude that the thermal shape fluctuations in medium-heavy nuclei should be important in a limited temperature range between $T\approx1$ MeV and $T\approx3$ MeV. They could be, perhaps, included on top of our model in the way proposed in Refs. \cite{OrmandBortignonBrogliaEtAl1990,OrmandBortignonBroglia1996}, however, we leave this beyond the scope of our paper admitting that thermal shape fluctuations have to be included for a correct comparison to experimental data.
%%%%%%%%%%%%
Apart from that, our results for $\Gamma(T)$ in the entire range of temperatures under study show a nearly quadratic dependence, which is in agreement with the Fermi liquid theory \cite{Landau1957}. The FT-RTBA results are also consistent with those of the microscopic approach of Ref. \cite{Bortignon1986}, which are available for the GDR energy region at $T\leq$ 3 MeV. 
%

\begin{figure*}
\begin{center}
\resizebox{0.90\textwidth}{!}{\includegraphics{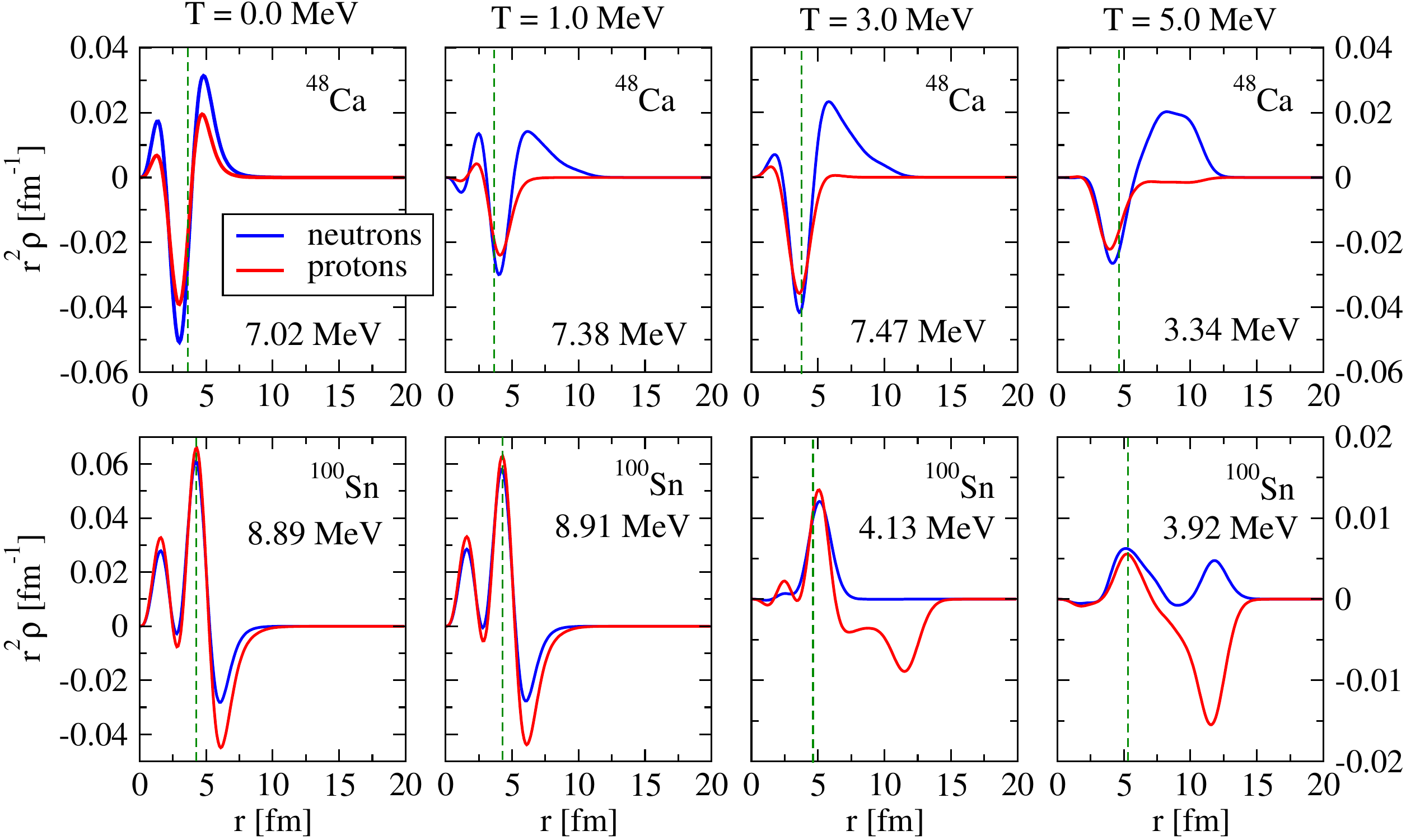}}
%\vspace{5cm}       % Give the correct figure height in cm
\end{center}
\caption{FT-RTBA transition densities of the strongest dipole states below 10 MeV in the neutron-rich $^{48}$Ca (top) and in the neutron-deficient $^{100}$Sn (bottom) at various temperatures. Vertical lines indicate the mean-field nuclear radius.}
\label{fig:8}       % Give a unique label
\end{figure*}
%
%
% BibTeX users please use

The energy-weighted sum rule for $^{48}$Ca  and $^{132}$Sn nuclei as a function of temperature within FT-RRPA and FT-RTBA is given in the right panel of Fig. \ref{fig:7}, in the percentage with respect to the Thomas-Reiche-Kuhn (TRK) sum rule. The EWSR at $T>0$ can be calculated in  analogy with the zero-temperature case \cite{Sommermann1983,Barranco1985a}. In our framework the meson-exchange interaction is velocity-dependent, so that already in RRPA and RQRPA at $T=0$ we observe up to  40\% enhancement of the 
%Thomas-Reiche-Kuhn (TRK) 
TRK sum rule within the energy regions which are typically studied in experiments \cite{LitvinovaRingTselyaev2007,LitvinovaRingTselyaev2008},  in agreement with data. In the resonant time blocking approximation without the ground state correlations associated with the PVC mechanism the EWSR is known to have exactly the same value as in RPA \cite{Tselyaev2007}. A small violation comes only with the subtraction procedure \cite{Tselyaev2007,LitvinovaTselyaev2007}. At $T=0$ in the RTBA corrected by the subtraction we find a few percent less EWSR in finite energy intervals below 25-30 MeV than in RRPA, however, this difference becomes smaller for larger intervals. As 
 in RTBA the strength distributions are more spread, one finds more strength outside finite intervals. 
The situation is very similar at T$>$0. In Fig. \ref{fig:7} (b) one can see that  the EWSR stays nearly flat with a slow decrease with the temperature growth due to the fact that the entire strength distribution shifts toward lower energies. The EWSR values from FT-RRPA and FT-RTBA  become nearly equal at $T=6$ MeV in both nuclei, because apparently at this temperature the high-energy tails of the strength distributions become less important.

As it follows from Eq. (\ref{Strength}), calculations of the nuclear response with small smearing parameter $\Delta$ are capable of resolving individual states in the low-energy region. Such calculations can reveal fine details of the evolution of the thermally emergent strength, as it was illustrated in Ref. \cite{WibowoLitvinova2018}. In addition, such calculations allow for extracting the transition densities from the response function, which provide a valuable information about the underlying structure of individual states.  Fig. \ref{fig:8} gives an illustration of such kind with the radial dependencies of the neutron and proton transition densities in the neutron-rich $^{48}$Ca (top panels) and in the neutron-deficient $^{100}$Sn (bottom panels). The proton and neutron transition densities are shown for the most prominent peaks below 10 MeV, which are selected separately for each temperature value. 
In $^{48}$Ca proton and neutron transition densities show in-phase oscillations inside the nucleus while neutron oscillations become weakly dominant outside at $T=0$. At $T=1$ MeV the major oscillations shift toward the nuclear surface while in the outer area one observes the typical pattern of the neutron skin oscillation known as pygmy dipole resonance. At $T=3$ MeV the general picture changes very little, but at $T=5$ MeV the oscillations of proton and neutron subsystems in the outer region start to develop the pattern of the out-of-phase oscillation. The latter is more typical for the giant dipole resonance and reproduced in numerous theoretical calculations at $T=0$. Our previous analysis of the thermal transition densities in a heavier nuclear system, such as $^{68}$Ni, has shown that the low-energy peak at high temperature has indeed some features of collective nature \cite{WibowoLitvinova2018}. 
%Thus, at very high temperature the coherent dipole oscillation shifts from the GDR region to the PDR region.  
In order to clarify whether this new feature is related to a considerable neutron excess, we have performed a similar analysis for the neutron-deficient $^{100}$Sn nucleus. As it can be seen from the bottom panels of Fig. \ref{fig:8}, at $T=0$, it also exhibits the in-phase oscillations of protons and neutrons inside the nucleus, but with a little dominance of proton oscillations in the surface area. Only little changes are observed at $T=1$ MeV, and at $T=3$ MeV one can see a clear proton dominance in the outer area. As in the case of $^{48}$Ca,  at $T = 5$ MeV one starts to distinguish a GDR-like pattern of the out-of-phase oscillation in the low-lying state at $E =$3.92 MeV while the in-phase oscillations are still present inside the nucleus and on the surface. Notice that  at $3\leq T \leq 5$ MeV the oscillations extend to far distances from the nuclear central region, which indicates that these oscillations are mainly associated with nucleons in the continuum. The latter is again consistent with the picture of Fig. \ref{fig:3} which illustrates  the possibility to create a large amount of $\widetilde{ph}$ transitions between the states located above $\varepsilon_F$.

\begin{figure}
%\vspace*{5cm}       % Give the correct figure height in cm
\resizebox{0.47\textwidth}{!}{\includegraphics{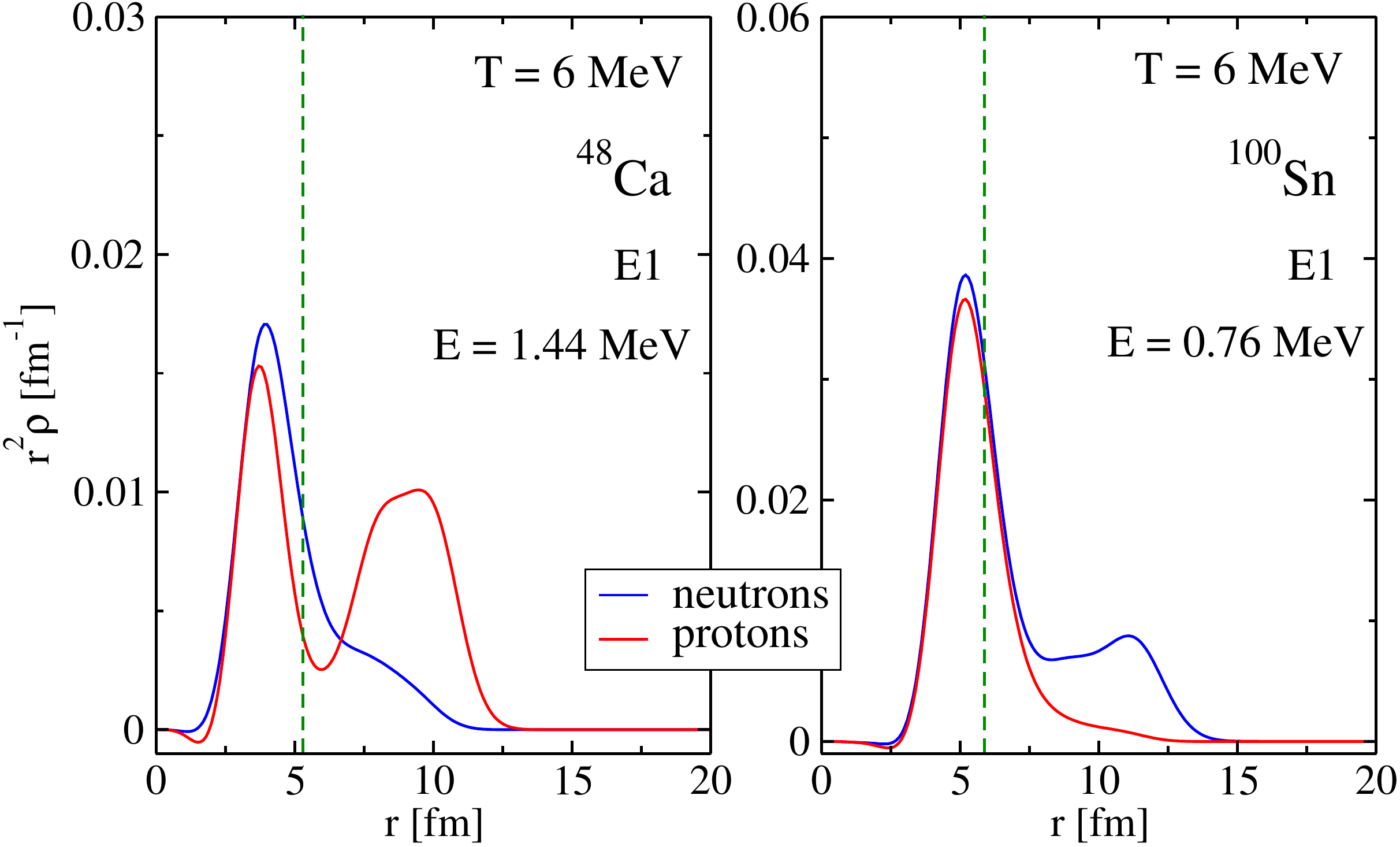}}
\caption{The proton and neutron FT-RRPA transition densities for the lowest dipole state at $T = 6$ MeV in $^{48}$Ca (left) and $^{100}$Sn (right).}
\label{fig:9}       % Give a unique label
\end{figure}

In the spectra of compound nuclei at finite temperature, especially at the high temperature values, the thermal unblocking allows for the appearance of the states with arbitrary small energy values. Such states can be indeed observed in Figs. \ref{fig:4} - \ref{fig:6} and, in principle, they would form a smooth continuum if the calculations were done in the complete basis \cite{LitvinovaBelov2013}. We have analyzed the underlying structure of the lowest dipole peaks at $E = $1.44 MeV in $^{48}$Ca and at $E = $0.76 MeV in $^{100}$Sn at $T=6$ MeV that is displayed in Fig. \ref{fig:9}. As at these transition energies the PVC effects included in the present model are rather weak, in this figure we show only the FT-RRPA calculations for the proton and neutron transition densities. In both neutron-rich $^{48}$Ca and neutron-deficient $^{100}$Sn we have obtained very similar pictures for the behavior of the proton and neutron transition densities in the lowest dipole states: they oscillate in phase and have their major peaks inside the nucleus while protons in $^{48}$Ca and neutrons in $^{100}$Sn show a pronounced minor peak in the outer area. The lowest dipole state at $E = $1.44 MeV in $^{48}$Ca is composed predominantly by the proton $\widetilde{ph}$ configuration $3p_{3/2}-2d_{5/2}$ and  the analogous state at $E = $0.76 MeV in $^{100}$Sn is dominated by the neutron $\widetilde{ph}$-configuration $1i_{13/2}-2h_{11/2}$. Other states in the lowest-energy domain are similar and their single  $\widetilde{ph}$-configuration character is in agreement with their relatively low transition probabilities.

%This picture reminds qualitatively the soft dipole mode in the two-neutron halo nucleus $^{11}$Li discussed in Ref. \cite{Broglia}

\section{Summary}
\label{summary}
In this article, which is dedicated to the memory and inspiring ideas of Pier Francesco Bortignon, we discuss a finite- temperature extension of the nuclear response theory beyond the relativistic RPA. We review the formalism of the thermal relativistic mean field and the time blocking method generalized for the finite-temperature case as the leading-order approach to the time-dependent part of the in-medium nucleon-nucleon interaction. Technically, the temperature-dependent imaginary-time projection operator is used to reduce the Bethe-Salpeter equation to an equation with one energy (frequency) variable in the energy domain. 

The  finite-temperature relativistic time blocking approximation was implemented on the base of the meson-nucleon Lagrangian of quantum hadrodynamics with the NL3 parametrization. We investigated the temperature dependence of the monopole, dipole and quadrupole response in 
the closed-shell medium-light $^{48}$Ca and of the dipole response of the medium-heavy $^{100,120,132}$Sn isotopes. It was found that the temperature dependence of the nuclear response is very generic and exhibits similar features for all multipoles and all nuclei under study. The most remarkable effects of the temperature increase on the spectra are (i)  the reinforcement of both the Landau damping and the particle-vibration coupling, (ii) the formation of the low-energy strength due to the thermal unblocking, and (iii) the shift of the entire strength distribution toward lower energies. We have investigated numerically the integral characteristics of the obtained dipole spectra, such as the energy-weighted sum rule and the width of the giant dipole resonance, in finite energy intervals up to high-temperature regimes. The EWSR was found to be very robust up to very high temperatures and the width exhibits a nearly parabolic growth with temperature, in accordance with the Landau theory. The obtained temperature dependence of the GDR's width is consistent with the available data, except for the temperature regime where thermal shape fluctuations become important. The results obtained for the dipole and quadrupole response functions are also consistent with the NFT calculations of Ref. \cite{Bortignon1986}. 

We have discussed some broader impacts of the presented developments, from the astrophysical r-process nucleosynthesis to the nuclear matter equation of state. The proposed treatment of the many-body correlations at finite temperature is of a general character and can be widely applied to the response of strongly-correlated systems other than finite nuclei. Future developments may consider the inclusion of continuum effects and ground state correlations associated with the particle-vibration coupling. 

%===============================================================================
\section*{Acknowledgements}

The authors greatly appreciate discussions with Peter Schuck and Jian Li. This work is supported by the NSF Career Grant PHY-1654379.
%
%===============================================================================

%\bibliographystyle{epj}
%\bibliography{Bibliography_Dec2018}
%\begin{thebibliography}{}
%$Author, \textit{Book title} (Publisher, place year) page numbers
% etc
%\end{thebibliography}

\end{document}